\documentclass[useAMS,usenatbib,usegraphicx]{mn2e}
\usepackage{bm}
\usepackage{times}

\title[Dust growth and grain size distribution]
{Effects of grain size distribution on the interstellar
dust mass growth}
\author[Hirashita \& Kuo]{Hiroyuki Hirashita$^1$\thanks{E-mail:
    hirashita@asiaa.sinica.edu.tw} and
    Tzu-Ming Kuo$^{1,2}$
\\
$^1$Institute of Astronomy and Astrophysics, Academia Sinica,
P.O. Box 23-141, Taipei 10617, Taiwan
\\ $^2$Department of Physics, National Taiwan University,
Taipei 10617, Taiwan
}
\date{2011 May 23}

\pagerange{\pageref{firstpage}--\pageref{lastpage}} \pubyear{2011}

\begin{document}
\label{firstpage}
\maketitle

\begin{abstract}
{Grain growth by the accretion of metals in interstellar
clouds (called `grain growth') could be} one of the
dominant processes that determine the dust content in
galaxies. The importance of grain size distribution for the
grain growth is demonstrated in this paper. First, we
derive an analytical formula that gives the grain size
distribution after the grain growth in
individual clouds for any initial grain size distribution.
The time-scale of the grain growth is very sensitive to
grain size distribution, since the grain growth is mainly
regulated by the surface to volume ratio of grains. Next,
we implement
the results of grain growth into dust enrichment models
of entire galactic system along with the
grain formation and destruction in the interstellar medium,
finding that the grain growth in
clouds governs the dust content in nearby galaxies
\textit{unless} the grain size is strongly biased to sizes
larger than $\sim 0.1~\micron$ or the power index
of the grain size distribution is shallower than
$\sim -2.5$.
The grain growth in clouds contributes to the rapid
increase of dust-to-gas ratio at a certain metallicity
level
{(called critical metallicity in \citealt{asano11}
and \citealt{inoue11}),} which we find to be
sensitive to grain size distribution.
Thus, the grain growth efficiently increase the dust
{mass} not only in nearby galaxies but also in
high-redshift quasars, whose metallicities are
larger than the critical value.
Our recipe for the grain growth is applicable
for any grain size distribution and easily implemented
into any framework of dust enrichment in galaxies.
\end{abstract}


\section{Introduction}

In the interstellar medium (ISM), dust grains are the
most efficient absorber of stellar light. The spectral
energy distributions and the radiative
heating and cooling in galaxies are 
thus strongly regulated by the presence of dust
\citep[e.g.][]{yamasawa11}. This means that the
understanding of dust enrichment in galaxies is
crucial in the studies
of galaxy evolution.

Dust enrichment in galaxies is governed by various
processes depending on age, metallicity, etc.\
\citep{dwek98}. In the earliest stage of galaxy
evolution, dust is predominantly produced by
supernovae (SNe) \citep[e.g.][]{kozasa09}, while at
later epochs asymptotic giant branch (AGB) stars
also contribute \citep{valiante09}. The time-scale
of dust destruction by SN shocks is
$\mbox{a few}\times 10^8$~yr
{\citep*{jones96,serra08}}, while that
of dust supply from stellar sources is longer than
1 Gyr in the Milky Way \citep{mckee89}.
Therefore, dust grains should grow in the ISM
by the accretion of metals onto grains \citep{draine09}
to explain the existence of dust in the ISM.
The growth occurs most efficiently in
molecular clouds, where the {typical}
number density of
hydrogen molecules is $\sim 10^3$ cm$^{-3}$
\citep{hirashita00}.
This process is called `grain growth in clouds' in this
paper.
Observational pieces of evidence for the grain growth
in clouds come from larger depletion of metal elements
in cold clouds than in warm medium
\citep{savage96}.

A lot of chemical evolution models treat the
evolution of dust content in galaxies. These models
usually include dust production by stars, grain
growth in clouds, and dust destruction by SNe
\citep*[e.g.][]{dwek98,inoue03,zhukovska08,calura08,
asano11}.
Most of the models which consider the grain growth in
clouds indicate that this process dominates the dust
budget at sub-solar or solar metallicities. The grain
growth occurs through the accretion of metals,
so that the increasing rate of grain mass by the
accretion of metals is proportional not only to the
metallicity but also to the grain surface-to-volume
ratio, which is very sensitive
to the grain size distribution.

Most models so far assume a certain grain size
distribution or a typical grain size to estimate the
surface-to-volume ratio. However, since the
dominant processes governing the grain size
distribution should vary with age and metallicity
\citep{odonnell97,hirashita10,yamasawa11},
it is expected that a variety of grain size distributions
emerge in a complex way depending on
age and metallicity. The first source of dust in the
history of galaxy
evolution is SNe, and the dust grains produced by
SNe are possibly biased to large
($\ga 0.1~\micron$) sizes because small grains
tend to be destroyed in the shocked region {within}
SNe {before being injected into the interstellar
space} \citep{bianchi07,nozawa07}. \citet{hirashita10}
show that small grains are produced by shattering
driven by interstellar turbulence if the dust abundance
is as high as that expected from the solar metallicity.
{\citet{jones96} show that shattering in interstellar
SN shocks increases the abundance of small grains.}
Efficient production of small grains enhances the
surface-to-volume ratio, activating the grain growth
by the accretion of metals. Thus, we should consider
various grain size distributions
depending on galaxy age and metallicity, and the
grain growth efficiency may vary with a
variety of grain size distributions.

The first aim of this paper is to formulate the grain
growth in clouds by explicitly considering the
dependence on grain size distribution. Therefore,
the former part of this
paper is devoted to the formulation of the grain
growth in clouds under an arbitrary grain size
distribution. Then, by using this formulation, we
point out the importance
of grain size distribution for the grain growth
{by accretion}.
The final scope of this work is to examine
if the grain size distribution has a significant
influence on
the dust enrichment in galaxies through the grain
growth in clouds.
Thus, in the latter part of this paper,
we implement our formulation of the grain growth
in clouds into a simple framework of dust
enrichment in
a galaxy, also taking into account the grain
formation by stellar sources and the
destruction by interstellar shocks driven by SN
remnants. Thereby, we will show
that our formulation of dust growth is
successfully incorporated into dust enrichment
models, and we will
address the
importance of grain size distribution for the
grain mass budget in galaxies.

This paper is organized as follows. We explain
the formulation in Section \ref{sec:formula}, and
describe some basic results on the evolution of
grain size
distribution through the grain growth in
individual clouds in
Section \ref{sec:result}. We implement the results
for the grain growth in clouds into a simple
evolution model of dust mass
in an entire galactic
system in Section \ref{sec:timescale},
where the model also treats the dust formation
by stellar sources and the dust destruction by
SN shocks.
We discuss the results in more general contexts in
Section \ref{sec:discussion}.
Finally, Section \ref{sec:conclusion} gives the
conclusion.

\section{Formulation}\label{sec:formula}

In this section, we formulate the evolution of
grain size distribution by the accretion of metals
(grain growth) in a single interstellar cloud.
Our procedures in this paper are divided into
the following two steps: (i) we construct a
formulation of the grain
growth in clouds, which is conveniently
incorporated in any dust evolution models in
galaxies; and (ii) we show that
our formulation in this section can be
used generally
to treat the grain growth in galaxies.
This section is aimed at item (i).
In Section \ref{sec:timescale}, we address 
item (ii)
by incorporating our formulation for the
grain growth into simple dust enrichment
models which also include dust formation by
stellar sources and dust destruction
in SN shocks. Other mechanisms that modify
the grain size distribution such as shattering
and coagulation are treated in other papers
\citep{jones94,jones96,yan04,hirashita09,yamasawa11}.
These processes are to be included in future work
for the comprehensive understanding of the
evolution of grain size distribution.

Throughout this paper, we call the elements
composing grains `metals'. We only treat grains
refractory enough to survive after the dispersal of
the cloud,
and do not consider volatile grains such as
water ice. More specifically, we consider silicate
and graphite as main dust components.
We also assume that the grains are spherical with
a constant material density $s$, so that the grain
mass $m$ and the grain radius $a$ are related as
\begin{eqnarray}
m=\frac{4}{3}\pi a^3s.\label{eq:mass}
\end{eqnarray}

\subsection{Evolution of grain size distribution}

We define the grain size distribution such that
$n(a,\, t)\,\mathrm{d}a$ is the number density of
grains whose radii are between $a$ and
$a+\mathrm{d}a$ at time $t$. For simplicity, we
assume that the gas density is constant
and the evolution of grain size distribution
occurs only through the accretion of metals
on dust grains.
In this situation,
the number density of grains is conserved. Thus, the
following continuity equation in terms of $n(a,\, t)$
holds:
\begin{eqnarray}
\frac{\partial n(a,\, t)}{\partial t}+
\frac{\partial}{\partial a}
\left[ n(a,\, t)\, u\right] =0,
\label{eq:continuity}
\end{eqnarray}
where $u\equiv\mathrm{d}a/\mathrm{d}t$
is the growth rate of the grain radius and
is given in the next subsection.
Coagulation and shattering, which do not
conserve the number density of grains, are not treated
to focus on the grain growth by accretion here.
Note that these processes do not change the grain
{mass}, while the grain growth by accretion
increases it. The evolution of
grain size distribution by coagulation and
shattering is treated in other papers
\citep{jones96,hirashita09,ormel09}.
We also neglect
possible grain destruction mechanisms in molecular
clouds by cosmic rays or shocks.

\subsection{Grain growth rate}\label{subsec:growth}

The grain growth rate is basically determined by
the collision rate between a grain and particles of
the relevant metal species. We adopt silicate
and graphite as dominant grain species
\citep[e.g.][]{draine84}, and denote elements
composing the grains as X (for example,
X = C, Si, etc.). We neglect the effect of Coulomb
interaction on the cross section (i.e.\ the cross
section of a grain for
accretion of metals is simply estimated by the geometric
one) because
the grains and the atoms are neutral in molecular
clouds \citep{weingartner99,yan04}.
{In fact, the ionization degree in dense clouds
is $10^{-6}$ \citep{yan04}, which means that almost
all the metal atoms colliding with the dust grains
are neutral.
}
The rate at which
atoms of element X strike the surface of a grain
with radius $a$ is denoted as $\mathcal{R}$ and is
estimated as \citep{evans94}
\begin{eqnarray}
\mathcal{R}=4\pi a^2n_\mathrm{X}\left(
\frac{kT_\mathrm{gas}}{2\pi m_\mathrm{X}}
\right)^{1/2},\label{eq:collision}
\end{eqnarray}
where $n_\mathrm{X}$ is the number density of
element X, $k$ is the Boltzmann constant,
$T_\mathrm{gas}$ is
the gas temperature, and
$m_\mathrm{X}$ is the atom mass of element X.
In general, dust grains are not composed of a
single species. We adopt
a key element, whose mass fraction in the grain
material is $f_\mathrm{X}$, and represent the
grain growth by the accretion of element X.
The concept of key element is also adopted by
\citet{zhukovska08}.

By using $f_\mathrm{X}$, the increase of the
grain mass $m$ is estimated as
\begin{eqnarray}
\frac{\mathrm{d}m}{\mathrm{d}t}=f_\mathrm{X}^{-1}
m_\mathrm{X}S\mathcal{R},\label{eq:dmdt}
\end{eqnarray}
where $S$ is the sticking probability. This
equation is converted into the
increasing rate of $a$ by using
equations (\ref{eq:mass}) and (\ref{eq:collision})
as
\begin{eqnarray}
u=\frac{\mathrm{d}a}{\mathrm{d}t}=
\frac{n_\mathrm{X}(t)\, m_\mathrm{X}S}
{f_\mathrm{X}s}
\left(\frac{kT_\mathrm{gas}}{2\pi m_\mathrm{X}}
\right)^{1/2}.\label{eq:dadt}
\end{eqnarray}
In fact, $n_\mathrm{X}(t)$ is a function of time
because the metal abundance in gas phase
decreases as the dust grains grow. This effect is
treated in the next subsection.

\subsection{Depletion of gas-phase metals}
\label{subsec:depletion}

The decreasing rate of the number density of
element X in gas phase is equal to the grain growth
rate per volume:
\begin{eqnarray}
\frac{\mathrm{d}n_\mathrm{X}}{\mathrm{d}t}=-
\int_0^\infty\mathcal{R}S\, n(a,\, t)\,\mathrm{d}a.
\label{eq:depletion}
\end{eqnarray}
Now we introduce the $\ell$-th moment of $a$ as
\begin{eqnarray}
\langle a^\ell\rangle (t)\equiv
\frac{1}{n_\mathrm{d}}
\int_0^\infty a^\ell n(a,\, t)\,\mathrm{d}a,
\label{eq:moment}
\end{eqnarray}
where $n_\mathrm{d}$ is the number density of
dust grains, which is independent of $t$:
\begin{eqnarray}
n_\mathrm{d}\equiv\int_0^\infty n(a,\, t)\,
\mathrm{d}a.\label{eq:n_d}
\end{eqnarray}
The moments are functions of $t$, and their values
at $t=0$ are denoted as $\langle a^\ell\rangle_0$
($\equiv\langle a^\ell\rangle (0)$).
By using the second moment of $a$ and
equation (\ref{eq:collision}) for $\mathcal{R}$,
equation (\ref{eq:depletion}) can be expressed as
\begin{eqnarray}
\frac{\mathrm{d}n_\mathrm{X}}{\mathrm{d}t}=-4\pi
n_\mathrm{X}(t)S\left(
\frac{kT_\mathrm{gas}}{2\pi m_\mathrm{X}}\right)^{1/2}
n_\mathrm{d}\,\langle a^2\rangle (t).\label{eq:dndt}
\end{eqnarray}

Note that the third moment of $a$ is related to
the dust mass density $\rho_\mathrm{d}(t)$ as
\begin{eqnarray}
\rho_\mathrm{d}(t)=\frac{4}{3}\pi\,
\langle a^3\rangle (t)\, sn_\mathrm{d}.
\label{eq:rho_d}
\end{eqnarray}
The initial dust mass density is
$\rho_\mathrm{d}(0)=\frac{4}{3}\pi\langle a^3
\rangle_0\, sn_\mathrm{d}$.

Here we quantify the initial number density of
element X.
The total number density of element X both in
gas and dust phases is written as
\begin{eqnarray}
n_\mathrm{X,tot}\equiv
\left(\frac{Z}{\mathrm{Z}_{\sun}}\right)
\left(\frac{\mathrm{X}}{\mathrm{H}}\right)_{\sun}
n_\mathrm{H},\label{eq:nxtot}
\end{eqnarray}
where $Z$ is the metallicity, and (X/H)$_{\sun}$ is
the solar abundance relative to hydrogen in
number, and $n_\mathrm{H}$ is the number density of
hydrogen nuclei. We denote the initial fraction of
element X in gas phase as $\xi$: 
\begin{eqnarray}
n_\mathrm{X}(0)=\xi\, n_\mathrm{X,tot}.\label{eq:zeta}
\end{eqnarray}
Since the initial number density
of element X in dust phase is
$f_X\rho_\mathrm{d}(0)/m_\mathrm{X}$, 
\begin{eqnarray}
\frac{f_\mathrm{X}\rho_\mathrm{d}(0)}{m_\mathrm{X}}
=(1-\xi )\, n_\mathrm{X,tot}.\label{eq:metallicity}
\end{eqnarray}
By using equation (\ref{eq:rho_d}) at $t=0$,
equation (\ref{eq:metallicity}) is written as
\begin{eqnarray}
1-\xi =
\frac{\frac{4}{3}\pi\langle a^3
\rangle_0f_\mathrm{X}sn_\mathrm{d}}
{m_\mathrm{X}n_\mathrm{X,tot}}.\label{eq:1-xi}
\end{eqnarray}
Thus, the normalization of the grain size distribution
is determined by
\begin{eqnarray}
n_\mathrm{d}=
\frac{m_\mathrm{X}(1-\xi )}
{\frac{4}{3}\pi\langle a^3\rangle_0f_\mathrm{X}s}
\left(\frac{Z}{\mathrm{Z}_{\sun}}\right)
\left(\frac{\mathrm{X}}{\mathrm{H}}\right)_{\sun}
n_\mathrm{H},\label{eq:nd}
\end{eqnarray}
where we used equation (\ref{eq:nxtot}) for
$n_\mathrm{X,tot}$.

\subsection{Formal solution}

The most important characteristics of the grain
growth by accretion is that the increasing rate of
$a$ is independent of $a$ (equation \ref{eq:dadt}).
Thus, equation (\ref{eq:continuity}) gives a formal
solution as
\begin{eqnarray}
n(a,\, t)=n(a-A(t),\, 0),\label{eq:formal}
\end{eqnarray}
where
\begin{eqnarray}
A(t)\equiv
\frac{m_\mathrm{X}S}{f_\mathrm{X}s}
\left(\frac{kT_\mathrm{gas}}{2\pi m_\mathrm{X}}
\right)^{1/2}\int_0^tn_\mathrm{X}(t)\,\mathrm{d}t.
\label{eq:A}
\end{eqnarray}
$A(t)$ is obtained if we give $n_\mathrm{X}(t)$
by using equation (\ref{eq:dndt}).
We formally assume that $n(a,\, 0)=0$
for $a<0$ so that equation (\ref{eq:formal})
can be used even for $a-A(t)<0$.

We introduce the following indicator for
the dust mass increase in the cloud
(equation \ref{eq:rho_d}):
\begin{eqnarray}
\frac{\rho_\mathrm{d}(t)}{\rho_\mathrm{d}(0)}
& = & \frac{\langle a^3\rangle (t)}{\langle a^3\rangle_0}
\nonumber\\
& = & \frac{\langle a^3\rangle_0+\langle a^2\rangle_0
A(t)+\langle a\rangle_0A(t)^2+A(t)^3}
{\langle a^3\rangle_0},
\label{eq:a3_expansion}
\end{eqnarray}
where we expand the third moment by using
equation (\ref{eq:formal}). Note that the dust mass
in the cloud at $t$ is
$\langle a^3\rangle (t)/\langle a^3\rangle_0$
times the initial dust mass.
{A similar equation is also obtained for the
grain mantle growth as shown by \citet*{guillet07}.}

\subsection{Typical time-scale}\label{subsec:norm}

For numerical calculations and interpretations of
the results, introducing a typical time-scale is
convenient. The typical time-scale of grain growth
by accretion is defined by
\begin{eqnarray}
\tau\equiv a_0\left/\left[
\frac{m_\mathrm{X}n_\mathrm{X,tot}S}{f_\mathrm{X}s}
\left(\frac{kT_\mathrm{gas}}{2\pi m_\mathrm{X}}
\right)^{1/2}\right]\right. ,\label{eq:tau}
\end{eqnarray}
where $a_0$ is a typical grain radius given
arbitrarily. 
By using $\tau$, equation (\ref{eq:A}) is reduced to
\begin{eqnarray}
A(t)=\frac{a_0}{\tau}\frac{1}{n_\mathrm{X,tot}}
\int_0^tn_\mathrm{X}(t)\,\mathrm{d}t,
\label{eq:A_final}
\end{eqnarray}
while equation (\ref{eq:dndt}) is written as
\begin{eqnarray}
\frac{\mathrm{d}n_\mathrm{X}(t)}{\mathrm{d}t}=-
\frac{3n_\mathrm{X}a_0}{\tau}(1-\xi )
\frac{\langle a^2\rangle (t)}{\langle a^3\rangle_0},
\label{eq:dndt_final}
\end{eqnarray}
where we have used equation (\ref{eq:1-xi}).
Note that
\begin{eqnarray}
\langle a^2\rangle (t)=\langle a^2\rangle_0+2
\langle a\rangle_0A(t)+A(t)^2.\label{eq:a2_expansion}
\end{eqnarray}
A set of
equations (\ref{eq:A_final})--(\ref{eq:a2_expansion})
is solved to
obtain the grain growth $A(t)$.

\subsection{Selection of quantities}
\label{subsec:quantities}

We consider silicate and graphite as representative
grain components \citep{draine84}.
In order to avoid the complexity arising from compound
species, we treat those two species separately as a
first approximation.
The quantities
adopted in this paper are summarized in
Table \ref{tab:material}.

\begin{table}
\centering
\begin{minipage}{80mm}
\caption{Adopted quantities.}
\label{tab:material}
    \begin{tabular}{lccccc}
     \hline
     Species & X & $f_\mathrm{X}$ & $m_\mathrm{X}$ [amu]
     & (X/H)$_{\sun}$ & $s$ [g cm$^{-3}$]\\ 
     \hline 
     Silicate & Si & 0.166 & 28.1 & $3.55\times 10^{-5}$ & 3.3 \\
     Graphite & C  & 1     & 12 & $3.63\times 10^{-4}$ & 2.26 \\
     \hline
    \end{tabular}
\end{minipage}
\end{table}

\citet{jones11} point out that the accretion of
Si, Fe, and Mg in the presence of abundant H$_2$,
CO and H$_2$O may lead to formation of a complex ice.
Therefore, the real picture of the silicate growth
in clouds may be chemically complicated.
They also state that any silicate materials produced
in the ISM with reasonable scenarios do not show
the spectral properties actually observed
in the ISM. For carbonaceous dust, graphite and
amorphous carbon have {different} chemical properties,
leading to different growth properties in accretion.
In this paper, because the knowledge about chemical
properties is still poor, we simplify the picture by
assuming that the grain growth is regulated
by the sticking of the key species.

\subsubsection{Silicate}

We assume that the accretion of silicon regulates
the growth of silicate grains. A larger abundance of
oxygen is compensated by a larger number required
to compose silicate.
\citet{zhukovska08} also consider Mg as well as
Si as a key species. The abundances of Mg and Si
are almost the same if we assume the solar
abundance pattern. Therefore, the following result
does not
change significantly even if we adopt O or Mg as
a key element.

The solar abundance of Si is
{
$\mathrm{(Si/H)}_{\sun}=4.07\times 10^{-5}$
\citep{lodders03}}, and
the mass of a Si atom is
$m_\mathrm{Si}=28.1$ amu
(1\,$\mathrm{amu}=1.66\times 10^{-24}$ g).
For the material properties of
silicate, we assume $s=3.3$ g cm$^{-3}$ and
$f_\mathrm{Si}=0.166$ based on composition
$\mathrm{Mg_{1.1}Fe_{0.9}SiO_4}$ \citep{draine84}.
By using
equation (\ref{eq:nxtot}), the typical
time-scale $\tau$ (equation \ref{eq:tau})
can be estimated as
\begin{eqnarray}
\tau =6.30\times 10^7a_{0.1}(Z/\mathrm{Z}_{\sun})^{-1}
n_3^{-1}T_{50}^{-1/2}S_\mathrm{0.3}^{-1}~\mathrm{yr},
\label{eq:tau_sil}
\end{eqnarray}
where $a_{0.1}\equiv a_0/0.1\,\micron$,
$n_3\equiv n_\mathrm{H}/10^3$\,cm$^{-3}$,
$T_{50}\equiv T_\mathrm{gas}/50$\,K, and
$S_\mathrm{0.3}=S/0.3$. Unless otherwise stated,
we adopt $n_\mathrm{H}=10^3$\,cm$^{-3}$ and
$T=50$\,K for the typical values derived from
observational properties of Galactic molecular
clouds \citep{hirashita00}, and $S=0.3$
\citep{leitch85,grassi11}.
Although $S$ may be almost 1 in such a
low temperature environments as molecular
clouds \citep{zhukovska08}, the chemical factor
(i.e.\ if the sticked
atom finally becomes the part of the grain or not)
is quite difficult to quantify as noted above
\citep{jones11}. Thus, we adopt a
conservative value of $S$ in this paper.
In fact, $\tau$ depends on $ZS$, so a larger value
of $S$ is compensated by a smaller $Z$. In other
words, the same dust growth rate is achieved
with metallicity $Z/(S/0.3)$ if we adopt a different
value of $S$.
Similar discussion also holds for the other quantities
($T$ and $n$).

\subsubsection{Graphite (carbonaceous grains)}

The solar abundance of C is
{
$\mathrm{(C/H)}_{\sun}=2.88\times 10^{-4}$
\citep{lodders03}}
and the mass of a C atom is $m_\mathrm{C}=12$ amu.
For graphite, we assume $s=2.26$ g cm$^{-3}$
\citep{draine84} and
$f_\mathrm{C}=1$ (i.e.\ graphite is solely composed
of C). The typical time-scale is
\begin{eqnarray}
\tau =5.59\times 10^7a_{0.1}
(Z/\mathrm{Z}_{\sun})^{-1}n_3^{-1}T_{50}^{-1/2}
S_{0.3}^{-1}~\mathrm{yr}.
\label{eq:tau_gra}
\end{eqnarray}
If we adopt other carbon species such as
amorphous carbon, the time-scale does not change
as long as $s$ is similar. If a light material like
hydrogenated amorphous
carbon (a-C:H) ($s=1.3$--1.5 g cm$^{-3}$ and
$f_\mathrm{C}=0.91$--0.98; \citealt{serra08}) is
adopted, $\tau$ becomes 0.52--0.65 times the
above value. The face value of $\tau$ indicates
that a-C:H grains grow more efficiently than
graphite, although the difference in $S$
between these two species is
not clear. {The destruction of
a-C:H is, however, also more efficient than
that of graphite \citep{serra08}.}

\subsection{Initial grain size distribution}
\label{subsec:initial}

We examine a variety of initial grain
size distributions.
The various models surveyed are called Models A--H
as listed in Table~\ref{tab:model}.

\subsubsection{Single size ($\delta$ function)}

We first examine the case where the grain size
distribution is strongly peaked at a certain
typical radius. For example, dust grains
condensed in SNe have a typical radius of
$\sim 0.1~\micron$ if the
shock efficiently destroys smaller grains
\citep{bianchi07,nozawa07}.
Here, the typical radius is
assumed to be $a_0$, and $\delta$
function is adopted for the grain size distribution
for simplicity. Then we obtain
\begin{eqnarray}
n(a,\, 0)=
n_\mathrm{d}\delta (a-a_0),
\end{eqnarray}
where $n_\mathrm{d}$ is given by
equation (\ref{eq:nd}).
The moments are
$\langle a^\ell\rangle_0=a_0^\ell$.

\subsubsection{Power law}\label{subsubsec:pwr}

If the grain size distribution is described by
a power law, the typical grain size is
not obvious. Thus, we arbitrarily adopt
$a_0=0.1~\micron$ for power-law grain
size distributions. The upper and lower
bounds for the grain radii are denoted as
$a_\mathrm{min}$ and
$a_\mathrm{max}$, respectively:
\begin{eqnarray}
n(a,\, 0)=\left\{
\begin{array}{ll}
{\displaystyle
\frac{(1-r)n_\mathrm{d}}
{(a_\mathrm{max}^{1-r}-a_\mathrm{min}^{1-r})}
}\, a^{-r}
& (r\neq 1); \\
{\displaystyle\frac{n_\mathrm{d}}
{\ln(a_\mathrm{max}/a_\mathrm{min})} }\,a^{-r}
& (r=1);
\end{array}
\right.
\end{eqnarray}
for $a_\mathrm{min}\leq{a}\leq a_\mathrm{max}$.
If $a<a_\mathrm{min}$ or $a>a_\mathrm{max}$,
$n(a,\, 0)=0$. The normalization $n_\mathrm{d}$
is given by equation (\ref{eq:nd}).

The moments are given by
\begin{eqnarray}
{\langle{a}^\ell\rangle_0} &
\hspace{-2mm}= & \hspace{-2mm}\left\{
\begin{array}{ll}
{\displaystyle\frac{1-r}{\ell +1-r}}\,
{\displaystyle
\frac{\, a_\mathrm{max}^{\ell +1-r}-
a_\mathrm{min}^{\ell +1-r}}
{a_\mathrm{max}^{1-r}-a_\mathrm{min}^{1-r}}}
& (r\neq 1,\, \ell +1);\\
{\displaystyle\frac{1}{\ell}}\,
{\displaystyle
\frac{\, a_\mathrm{max}^{\ell}-a_\mathrm{min}^{\ell}}
{\ln (q/p)}} &
(r=1);\\
{\displaystyle\frac{-\ell}
{a_\mathrm{max}^{-\ell}-a_\mathrm{min}^{-\ell}}}
\ln(a_\mathrm{max}/a_\mathrm{min})
& (r=\ell +1).\\
\end{array}
\right.\nonumber\\ & &
\end{eqnarray}

\section{Grain growth in individual clouds}
\label{sec:result}

\begin{table*}
\centering
\begin{minipage}{130mm}
\caption{Models.}
\label{tab:model}
    \begin{tabular}{ccccccccccc}
     \hline
     Model & ${n}\,^\mathrm{a}$ &
     $a_0$ & $\xi$ & $r$ & $a_\mathrm{min}$ &
     $a_\mathrm{max}$ &
     $\langle{a}\rangle_0\,^\mathrm{c}$ &
     $\sqrt{\langle{a}^2\rangle_0}\,^\mathrm{c}$ &
     $\sqrt[3]{\langle{a}^3\rangle_0}\,^\mathrm{c}$ &
     $\langle a^3\rangle_0/\langle a^2\rangle_0\,^\mathrm{c}$\\
      & & [$\micron$] & & & [$\micron$] & [$\micron$] &
      [$\micron$] & [$\micron$] & [$\micron$] &
      [$\micron$] \\
     \hline 
     A & $\delta$ & ---$^\mathrm{b}$ & 0.5 & --- & ---
       & --- & --- & --- & --- & ---\\
     B & $\delta$ & ---$^\mathrm{b}$ & 0.9 & --- & ---
       & --- & --- & --- & --- & ---\\
     C & p & 0.1 & 0.5 & 3.5 & 0.001 & 0.25 & 0.00167 &
       0.00216 & 0.00402 & 0.0159\\
     D & p & 0.1 & 0.5 & 2.5 & 0.001 & 0.25 & 0.00281 &
       0.00667 & 0.0158 & 0.0887\\
     E & p & 0.1 & 0.5 & 4.5 & 0.001 & 0.25 & 0.00140 &
       0.00153 & 0.00187 & 0.00279\\
     F & p & 0.1 & 0.9 & 3.5 & 0.001 & 0.25 & 0.00167 &
       0.00216 & 0.00420 & 0.0159\\
     G & p & 0.1 & 0.5 & 3.5 & 0.0003 & 0.25 & 0.000499 &
       0.000659 & 0.00156 & 0.00874\\
     H & p & 0.1 & 0.9 & 3.5 & 0.003 & 0.25 & 0.00499 &
       0.00633 & 0.0103 & 0.0273\\
     \hline
    \end{tabular}

\medskip

$^\mathrm{a}$ Initial grain size distribution: ``p'' for
the power law and ``$\delta$'' for the $\delta$ function.
\\
$^\mathrm{b}$ The time-scale is scalable for any $a_0$
by using equations (\ref{eq:tau_sil}) and
(\ref{eq:tau_gra}).
\\
$^\mathrm{c}$ The moments of ${a}$ are not free
parameters but are calculated once the other
parameters are fixed.
\end{minipage}
\end{table*}

\begin{figure}
\includegraphics[width=0.45\textwidth]{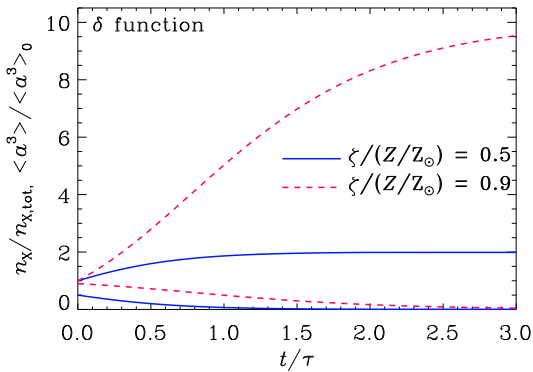}
 \caption{Evolutions of the normalized grain volume
 (or mass) relative to the initial value,
 $\langle a^3\rangle /\langle a^3\rangle_0$,
 and the number density of element X in gas phase
 normalized to the total (gas phase + dust phase)
 number density, ${n}_\mathrm{X}/n_\mathrm{X,tot}$,
 along with the normalized time ${t}/\tau$.
 The $\delta$ function grain size distribution is
 adopted. Solid and dashed lines
 show the results of Models A and B, respectively
 (i.e.\ $\xi =0.5$ and 0.9,
 respectively) with the upper and lower lines show the
 evolution of
 $\langle{a}^3\rangle /\langle{a}^3\rangle_0$
 and ${n}_\mathrm{X}/n_\mathrm{X,tot}$, respectively.
 }
 \label{fig:delta}
\end{figure}

First, we examine single size cases (Models A and B),
where the grain size distribution is described by
$\delta$ function. For the initial fraction of element X
in gas phase, we adopt $\xi =0.5$ and 0.9 as
representative cases where a moderate fraction and
only a small fraction of metals are in dust phase,
respectively. In Fig.\ \ref{fig:delta},
we show the evolution of
${n}_\mathrm{X}(t)/n_\mathrm{X,tot}$ and
$\langle{a}^3\rangle (t)/\langle{a}^3\rangle_0$
for Models A and B. Note that
$\langle{a}^3\rangle (t)/\langle{a}^3\rangle_0$
is proportional to the dust mass or the dust mass
density (equation \ref{eq:a3_expansion}) and that
${n}_\mathrm{X}(t=0)/n_\mathrm{X,tot}=\xi$
(the fraction of metals in gas phase at $t=0$).
As the model imposes, the dust mass increases
with the depletion of
gas-phase metals onto the grains.
For larger $\xi$, the grains continue
to grow to larger radii and for a longer time,
since the available metals
in gas phase is more abundant and
the abundance of dust relative to the gas phase
metals is smaller. At ${t}\gg\tau$,
$\langle{a}^3\rangle (t)/\langle{a}^3\rangle_0$
approaches $1/(1-\xi )$, which means that the
grain mass becomes $1/(1-\xi )$ times the initial
mass after all the gas phase X is depleted onto
the dust grains.

Next, we examine the power-law grain size
distributions (Models C--H in Table \ref{tab:model}).
\citet[][hereafter MRN]{mathis77} show that the
extinction curve in the Milky Way can be fitted
with a power-law grain size distribution.
We fix the range of the grain size by adopting
$a_\mathrm{min}=0.001~\micron$, and
$a_\mathrm{max}=0.25~\micron$ (MRN).
We arbitrarily fix the value of $a_0$ as
$a_0=0.1~\micron$ (Section \ref{subsubsec:pwr}).
Since the lower bound of the
grain size is poorly determined from the extinction
curve \citep{weingartner01},
we also examine the case
where the smallest
radius is 0.0003 $\micron$ (3 \AA).
We change the power-law index $r$ from 2.5 to 4.5
to examine the cases where large and small grains are
relatively dominated, respectively (note that $r=3.5$
corresponds to MRN). 
We also vary $\xi$.

\begin{figure}
\includegraphics[width=0.45\textwidth]{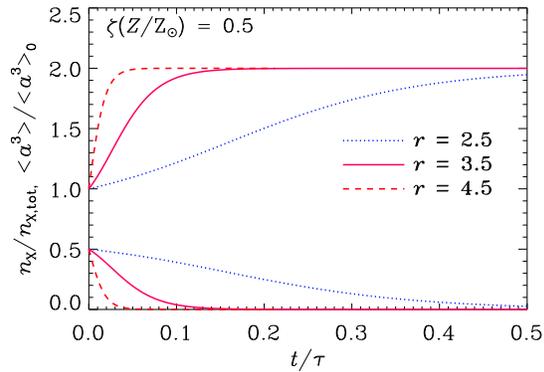}
 \caption{Same as Fig.\ \ref{fig:delta} but for the
 power-law initial grain size distributions (Models
 C--E). Solid, dotted, and dashed lines
 show the results of Models C, D, and E ($r=3.5$,
 2.5, and 4.5), respectively
 with the upper and lower lines show the
 evolution of
 $\langle{a}^3\rangle /\langle{a}^3\rangle_0$
 and
 ${n}_\mathrm{X}/n_\mathrm{X,tot}$, respectively.
 }
 \label{fig:pwr}
\end{figure}

In Fig.\ \ref{fig:pwr}, we show the results for
Models C--E.
As shown in equation (\ref{eq:dndt_final}), the
consumption rate of metals in gas phase is
proportional to the
surface-to-volume ratio of dust grains. If $r$ is large,
the surface-to-volume ratio is large,
so that the metals in gas phase is rapidly consumed onto
dust grains.

\begin{figure}
\includegraphics[width=0.45\textwidth]{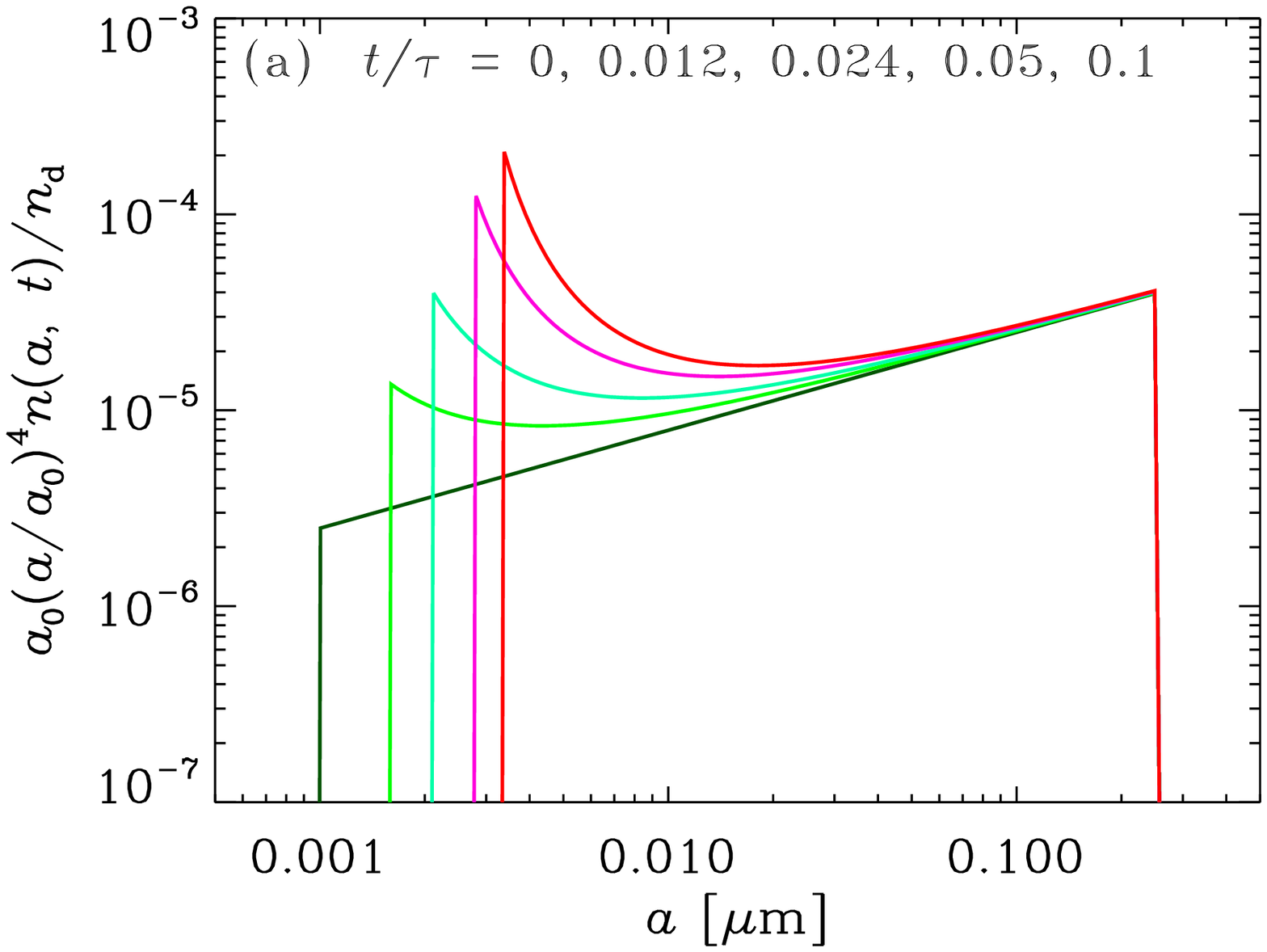}
\includegraphics[width=0.45\textwidth]{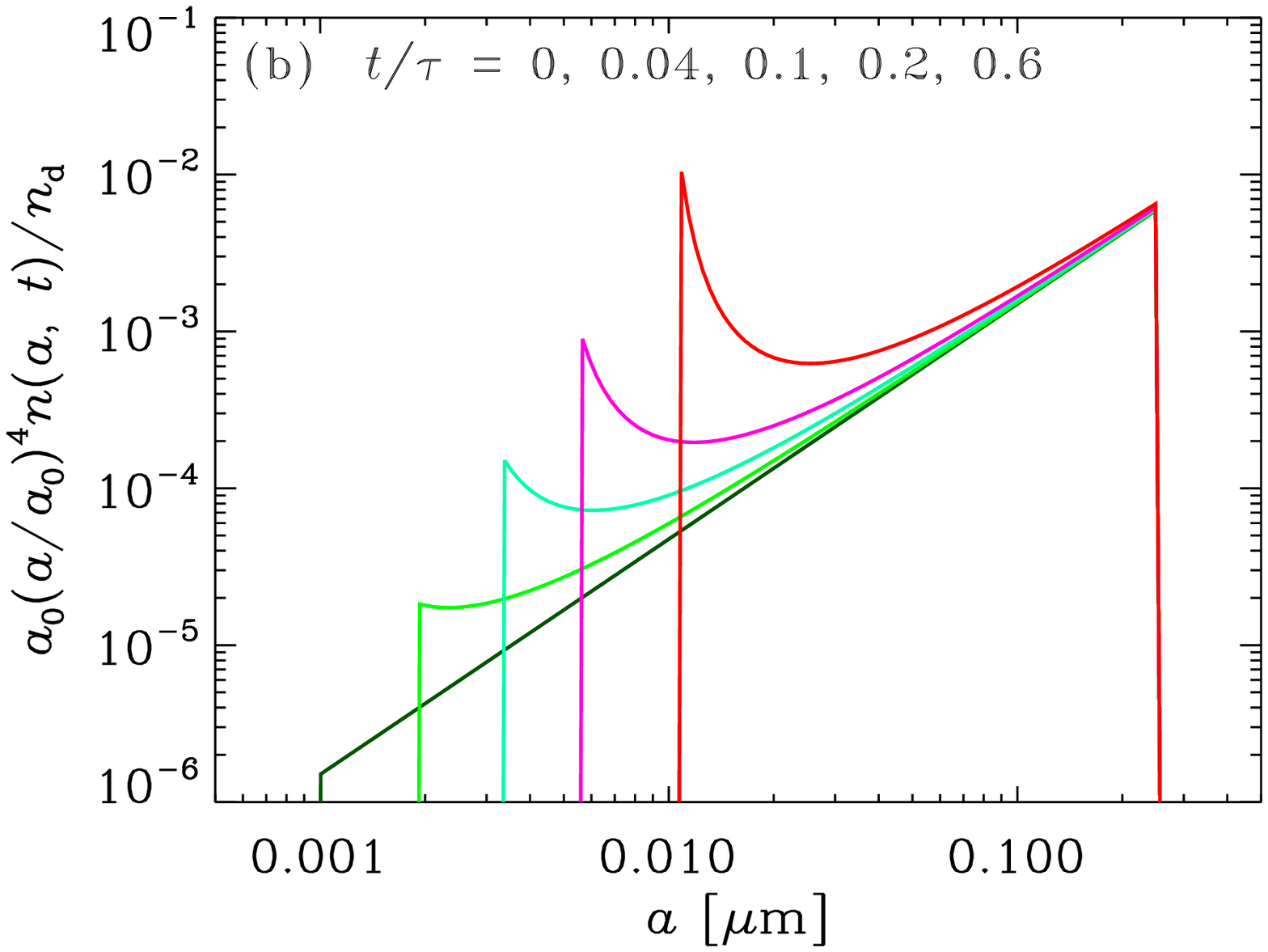}
\includegraphics[width=0.45\textwidth]{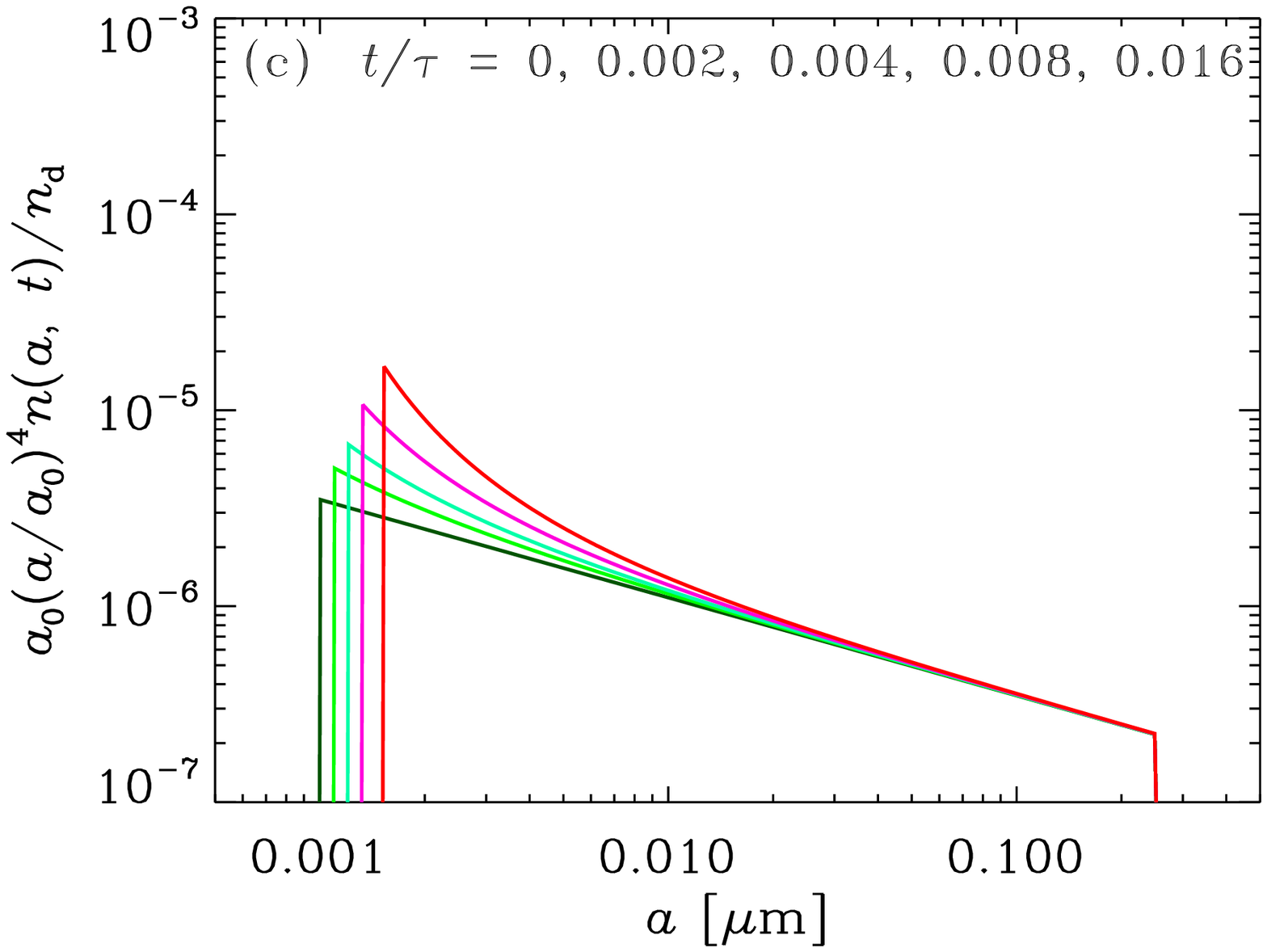}
 \caption{Evolution of grain size distribution for
 Models C, D, and E (Panels a, b, and c, respectively).
 The normalized times are shown in each panel. The
 size distribution is shown by
 $a_0(a/a_0)^4{n}({a},\,{t})/n_\mathrm{d}$
 to indicate the normalized grain mass distribution
 per logarithmic size.
 }
 \label{fig:size_ev}
\end{figure}

Fig.\ \ref{fig:size_ev} illustrates the evolution of
grain size distribution. To show the mass distribution
per logarithmic size, we multiply
${a}^4$ to ${n}$. Also in order to make the quantity
dimensionless, we multiply $a_0/n_\mathrm{d}$. We indeed
observe that the small grains grow to larger grains.
Since the increasing
rate of grain radius is independent of $a$, the
impact of grain growth is significant for small
grains {as already shown by \citet{guillet07}
for the mantle growth}.
Moreover, gas-phase metals accrete
selectively onto small grains
{(especially for larger $r$)} because the grain surface
is dominated by small grains;
{this is consistent with the
results of \citet{weingartner99}}.

\begin{figure}
\includegraphics[width=0.45\textwidth]{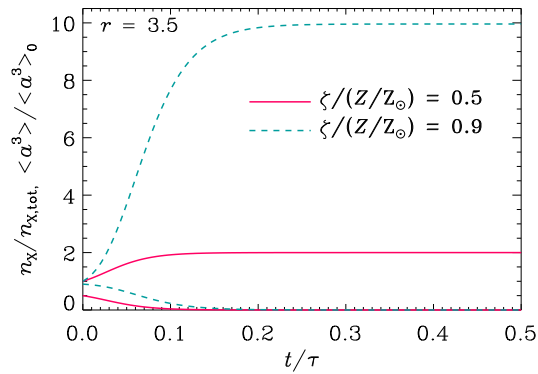}
 \caption{Same as Fig. \ref{fig:pwr} but for
 Models C and F
 ($\xi =0.5$ and 0.9, respectively, with $r=3.5$).
 }
 \label{fig:pwr_depl}
\end{figure}

To show the dependence on $\xi$, we compare
Models C and F. The results are shown in
Fig.\ \ref{fig:pwr_depl}. Different values of $\xi$
lead to different final-to-initial dust mass ratios
$\langle{a}^3\rangle /\langle{a}^3\rangle_0\to 1/(1-\xi )$,
which are interpreted in the same way as in the $\delta$
function cases (Fig.\ \ref{fig:delta}).

\begin{figure}
\includegraphics[width=0.45\textwidth]{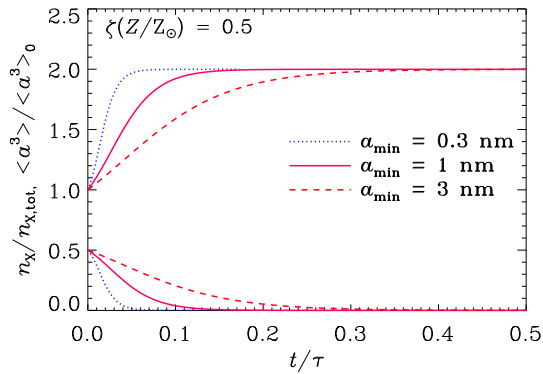}
 \caption{Same as Fig. \ref{fig:pwr} but for
 Models C, G, and H, where the minimum grain radius is
 varied as $a_\mathrm{min}=0.3$, 1, and 3 nm,\
 respectively, with $r=3.5$.
 }
 \label{fig:pwr_amin}
\end{figure}

Finally, in Fig.\ \ref{fig:pwr_amin}, we compare
Models C, G, and H to examine the dependence
on $a_\mathrm{min}$. As expected, the
results are sensitive to $a_\mathrm{min}$ since
the surface-to-volume ratio of grains differs largely
(Table \ref{tab:model})
\citep[see also][]{weingartner99}.
Therefore, it is important
to specify the physics governing the smallest grain
size in interstellar clouds. Coagulation possibly
depletes grains smaller than
{several \AA\ \citep{hirashita09,guillet11}}.

\section{Implementation into galaxy evolution models}
\label{sec:timescale}

In this section, we relate the results above for
the grain growth in individual clouds to
the dust mass evolution in an entire galactic
system. We adopt a chemical evolution model
that calculates the metal and dust enrichment.
To concentrate on the importance of grain
size distribution, we simplify the galaxy
evolution models, although our recipe for the
grain growth is simple enough to be easily
implemented to more complicated galaxy evolution
models. The entire galactic system is
treated as one zone. {The treatment of
one zone is good if}
the dust, metals, and
gas are mixed instantaneously {or if we
focus on a certain well mixed region in the
galaxy. For example, in spiral galaxies which
generally have a radial metallicity gradient,
our models can be applicable to a certain radius
range where the metallicity can be regarded as
uniform.}

Since we are interested in the effects on
dust enrichment, we compare our results with the
dust abundance in galaxies. Given that the dust
enrichment is closely related with the metal
enrichment, it
is convenient to derive the relation between
dust-to-gas ratio and metallicity
\citep[e.g.][]{lisenfeld98}. The dust mass is
usually estimated by the far-infrared emission
to trace the emission from large grains, which
occupy a significant fraction of the dust
mass. The dust mass estimated from the
far-infrared emission can miss very small
grains and very cold dust.
The contribution from very small
grains to the dust mass is not significant
{\citep{desert90,galliano05,compiegne11}}. Very
cold dust traced
in longer wavelengths than submillimetre
may have a significant contribution to the
total dust mass especially in metal-poor
galaxies \citep{galliano03,galliano05,galametz11},
but its abundance is
significantly affected by the assumed
emissivity index of large grains.
If we miss the contribution from very cold dust,
the observational dust-to-gas ratio is
underestimated in this paper, which
enhances the importance of the dust growth
in clouds to explain the additional contribution
from very cold dust.

Although there is a large variety in the
grain size distribution derived from the
dust emission spectrum \citep{galliano05},
the regulation mechanism of grain size
distribution in the ISM is not fully understood.
Therefore, in this paper, we examine various
grain size distributions (i.e.\ $\delta$ function
with various typical sizes and power law with
various power indeces). Sputtering
and shattering in
SN shocks \citep{jones96} and shattering in
interstellar turbulence \citep*{yan04,hirashita10}
may play a significant role in determining the
shape of grain size distribution. Inclusion of
these processes into the evolution of grain
size distribution is left for future work to
concentrate on the grain growth in clouds
in this paper. A simultaneous treatment of
dust formation and destruction
can be seen in \citet{yamasawa11}.

Photo-processes are neglected in this paper.
Dust may contain ice mantle when it is injected
into the ISM from clouds. When the dust is
exposed to stellar radiation, the evaporation of
the ice mantle may lead to grain disaggregation.
If small refractory grains are ejected in the
disaggregation, the total amount of refractory
dust (silicate and graphite in this paper) does not
change, so that our conclusion below is not affected.
If the dust is destroyed and the refractory elements
are returned into the gas phase in the
photo-processing, the dust abundance may be
affected by this process. The destruction
of dust by photo-processing, if any, can be
effectively included into the dust destruction
efficiency ($\beta_\mathrm{SN}$ defined later),
but we will see later that only the destruction
by SNe is enough to explain the dust-to-metal
ratio (Section \ref{subsec:Zcr}).

\subsection{Grain growth time-scale}
\label{subsec:dMdt1}

We define the increased fraction of dust mass in
a cloud, $\beta$, which is evaluated as
\begin{eqnarray}
\beta & = &
\frac{\langle a^3\rangle (\tau_\mathrm{cl})}
{\langle a^3\rangle_0}-1\nonumber\\
& = &
\frac{A(\tau_\mathrm{cl})[3\langle a^2\rangle_0
+3\langle a\rangle_0
A(\tau_\mathrm{cl})+A(\tau_\mathrm{cl})^2]}
{\langle a^3\rangle_0},\label{eq:beta}
\end{eqnarray}
where $\tau_\mathrm{cl}$ is the lifetime of
clouds hosting the grain growth and
equation (\ref{eq:a3_expansion}) is used
from the second to the third step. The dust mass
in the cloud becomes $(\beta +1)$ times the initial
value after the cloud lifetime, when the dust
grown in the cloud returns in the diffuse ISM
(see equation \ref{eq:a3_expansion}).

Now we consider how to implement our results
into dust enrichment models for an entire
galactic system. By using $\beta$, the
increasing rate of dust mass
by accretion in clouds should be written as
\begin{eqnarray}
\left[\frac{\mathrm{d}M_\mathrm{dust}}{\mathrm{d}t}
\right]_\mathrm{acc}=
\frac{\beta X_\mathrm{cl}M_\mathrm{dust}}
{\tau_\mathrm{cl}},\label{eq:dMdt2}
\end{eqnarray}
where $M_\mathrm{dust}$ is the total dust mass 
in the galaxy, and $X_\mathrm{cl}$ is the mass
fraction of clouds hosting the grain growth to the
total gas mass. In equation (\ref{eq:dMdt2}), we
have assumed that the time-scale of
dust enrichment is much longer than the cloud
lifetime so that the mass increasing rate in
each cloud is estimated by
$\beta /\tau_\mathrm{cl}$ times the dust mass
in the cloud. On the other hand,
\citet{hirashita00} express the increasing
rate of the dust mass in a galaxy through the
grain growth in clouds by
\begin{eqnarray}
\left[
\frac{\mathrm{d}M_\mathrm{dust}}{\mathrm{d}t}
\right]_\mathrm{acc}=
\frac{X_\mathrm{cl}M_\mathrm{dust}\xi}
{\tau_\mathrm{grow}},\label{eq:dMdt1}
\end{eqnarray}
where
$\tau_\mathrm{grow}$ is the
growth time-scale of the dust in a cloud, and
$\xi$ is the fraction of metals in gas phase
(Section \ref{subsec:growth}).
A similar expression for the increasing rate of
the dust mass is widely adopted in dust enrichment
models \citep[e.g.][]{dwek98,inoue03,calura08}.

\begin{figure}
\includegraphics[width=0.45\textwidth]{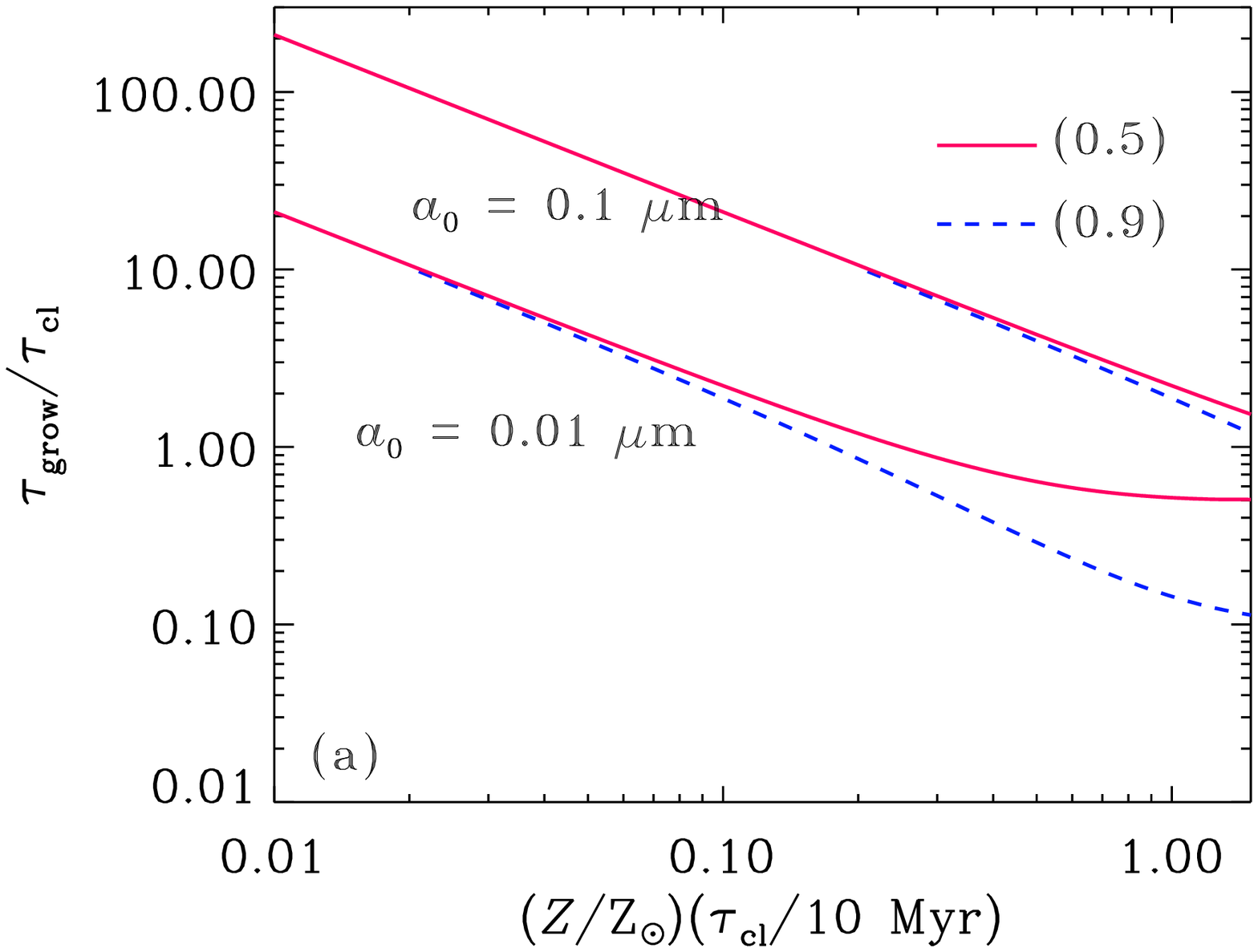}
\includegraphics[width=0.45\textwidth]{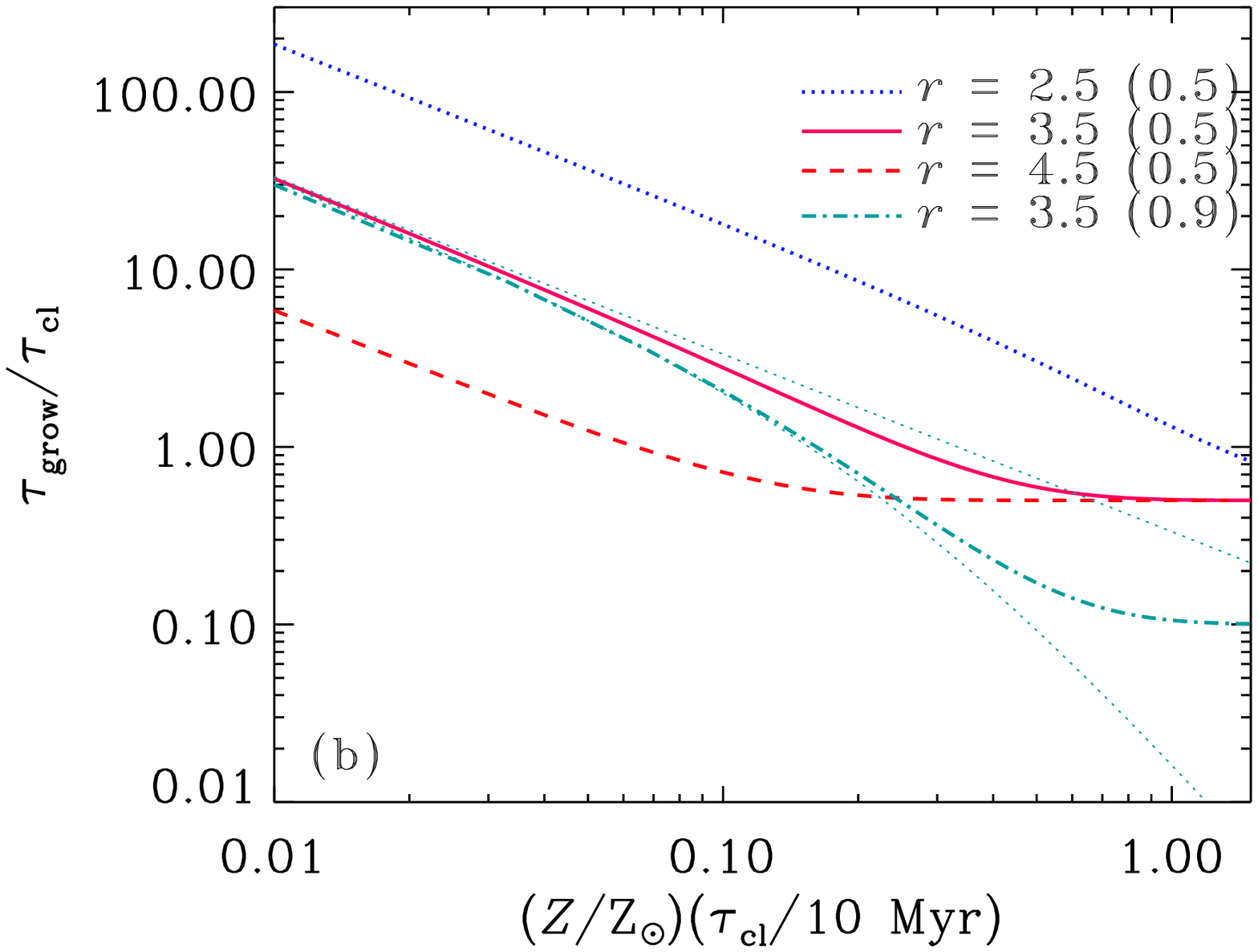}
 \caption{Ratio of the grain growth time-scale to the
 lifetime of clouds hosting the grain growth by accretion
 ($\tau_\mathrm{grow}/\tau_\mathrm{cl}$) as a function of
 $Z\tau_\mathrm{cl}$. (a) The cases of single grain sizes
 ($\delta$ function size distributions) are shown.
 The solid and dashed lines show the cases for
 $\xi =0.5$ and 0.9, respectively. The
 upper and lower lines show the cases for
 $a_\mathrm{0}=0.1$ and 0.01 $\micron$, respectively.
 (b) The results for power-law grain size distributions are
 presented. The thick dotted, solid, and dashed lines are for
 $r=2.5$, 3.5, and 4.5 with $\xi =0.5$,
 while the dot-dashed line shows the case for $r=3.5$ with
 $\xi =0.9$. The thin dotted lines
 show the relation predicted from
 equation (\ref{eq:taugrow_smallZ}) (upper line) and
 equation (\ref{eq:taugrow_taucl}) with
 equation (\ref{eq:A_lowZ})
 (lower line) (the dot-dashed line is relevant for
 comparison).
 }
 \label{fig:taugrow}
\end{figure}

Comparing equations (\ref{eq:dMdt2}) and
(\ref{eq:dMdt1}), we obtain
\begin{eqnarray}
\frac{\tau_\mathrm{grow}}{\tau_\mathrm{cl}}=
\frac{\xi}{\beta}=
\frac{\xi\langle a^3\rangle_0}
{A(\tau_\mathrm{cl})[3\langle a^2\rangle_0
+3\langle a\rangle_0
A(\tau_\mathrm{cl})+A(\tau_\mathrm{cl})^2]},
\label{eq:taugrow_taucl}
\end{eqnarray}
where equation (\ref{eq:beta}) is used for $\beta$.
Thus, we obtain
$\tau_\mathrm{grow}/\tau_\mathrm{cl}$ as
a function of $\tau_\mathrm{cl}$; that is, if we
give the lifetime of molecular cloud, we can
evaluate $\tau_\mathrm{grow}$.
Equation (\ref{eq:A_final}) indicates that
$A(\tau_\mathrm{cl})$ depends on $\tau$. Since $\tau$
depends on the metallicity
(equations \ref{eq:tau_sil} and \ref{eq:tau_gra}),
$\tau_\mathrm{grow}/\tau_\mathrm{cl}$ evolves
as the system is enriched with metals. Here we adopt
equation (\ref{eq:tau_sil}) for $\tau$. To simplify
the discussion, we focus on the metallicity and
the grain size distribution with the other quantities
fixed; that is,
$\tau =63.0a_\mathrm{0.1}(Z/\mathrm{Z}_{\sun})^{-1}$ Myr.
As a result, if we specify $\tau_\mathrm{cl}$ and give a
grain size distribution, we obtain
$\tau_\mathrm{grow}/\tau_\mathrm{cl}$ as a function of
$Z\tau_\mathrm{cl}$. Thus, we can regard
$\tau_\mathrm{grow}/\tau_\mathrm{cl}$ as a function
of metallicity if we fix $\tau_\mathrm{cl}$.
The results are shown in Fig.\ \ref{fig:taugrow}.

Fig.\ \ref{fig:taugrow} shows that
$\tau_\mathrm{grow}/\tau_\mathrm{cl}$ depends on
the grain size distribution as expected from the
results in the previous section. In particular,
if $a_0$ is smaller for the $\delta$ function
cases or $r$ is larger in the power-law cases,
the grains grow on a shorter time-scale so
$\tau_\mathrm{grow}$ becomes smaller. We also
observe the dependence on $\xi$: if $\xi$ is larger,
the grains grow to a larger
extent because the abundance of available metals
relative to dust is larger. Thus,
$\tau_\mathrm{grow}/\tau_\mathrm{cl}$ becomes
smaller for larger $\xi$ in the region of high
$Z$ or large $\tau_\mathrm{cl}$.

Fig.\ \ref{fig:taugrow} also shows that the
behaviour at small
metallicities is described by
$\tau_\mathrm{grow}\propto Z^{-1}$. 
This is explained
as follows. If the metallicity is low enough,
$\tau_\mathrm{cl}\ll\tau$.
Thus, the
decrease of gas-phase metals is negligible,
and equation (\ref{eq:A_final}) can be written
as follows if we recall that
${n}_\mathrm{X}(0)/n_\mathrm{X,tot}=\xi$:
\begin{eqnarray}
{A}(\tau_\mathrm{cl})\simeq\xi\,
a_0\tau_\mathrm{cl}/\tau~~~
\mbox{for $\tau_\mathrm{cl}\ll\tau$}.\label{eq:A_lowZ}
\end{eqnarray}
By taking up to
the first order for $\tau_\mathrm{cl}/\tau$,
equation (\ref{eq:taugrow_taucl}) is
approximated as
\begin{eqnarray}
\tau_\mathrm{grow}\simeq
\frac{\langle{a}^3\rangle_0}{3\langle{a}^2\rangle_0\, a_0}
\tau\propto Z^{-1}~~~\mbox{for $\tau_\mathrm{cl}\ll\tau$}.
\label{eq:taugrow_smallZ}
\end{eqnarray}
This indicates two important properties. One is
that the grain growth time-scale (which is
proportional to $Z^{-1}$) is roughly determined by
the typical growth time-scale $\tau$ and the
surface-to-volume ratio of dust grains. The other
is that the cloud lifetime does not enter since the
grain growth in each cloud during the cloud lifetime
is slight enough.

As $Z$ becomes large,
$\tau_\mathrm{grow}/\tau_\mathrm{cl}$ approaches
a constant value independent of metallicity. Indeed,
the grain growth saturates if the metals in gas phase
are used out. Thus,
$\langle{a}^3\rangle (\tau_\mathrm{cl})/
\langle{a}^3\rangle_0\to 1/(1-\xi )$
as $Z$ becomes large.
In reality, the grain abundance in clouds would
approach the value
expected from the equilibrium between the grain
growth and the destruction by cosmic ray, shocks,
ultraviolet light in clouds, etc. We expect that the
destruction within the clouds does not
affect the dust abundance in the entire galaxy
since the dominant destruction occurs in
the diffuse ISM by SN shocks \citep{mckee89}.
Thus, in the clouds, we only consider the grain
growth and neglect dust destruction. In the
dust enrichment
model for the entire galaxy in
Section \ref{subsec:enrichment},
we include the dust destruction by SN
shocks in diffuse medium.
By using
equation (\ref{eq:taugrow_taucl}), we obtain
\begin{eqnarray}
\tau_\mathrm{grow}\simeq
(1-\xi )\tau_\mathrm{cl}~~~
\mbox{(for $\tau_\mathrm{cl}\gg\tau$)},
\end{eqnarray}
where we note that $\beta\simeq\xi /(1-\xi )$
for $\tau_\mathrm{cl}\gg\tau$.
This result indicates that the grain growth is
regulated by the cloud lifetime since the dust grains
use up
all the gas-phase metals within
$\tau_\mathrm{cl}$.
The factor $(1-\xi)$ indicates that the dust mass
increases with a larger fraction if a larger part
of metals are in gas phase.

\begin{figure}
\includegraphics[width=0.45\textwidth]{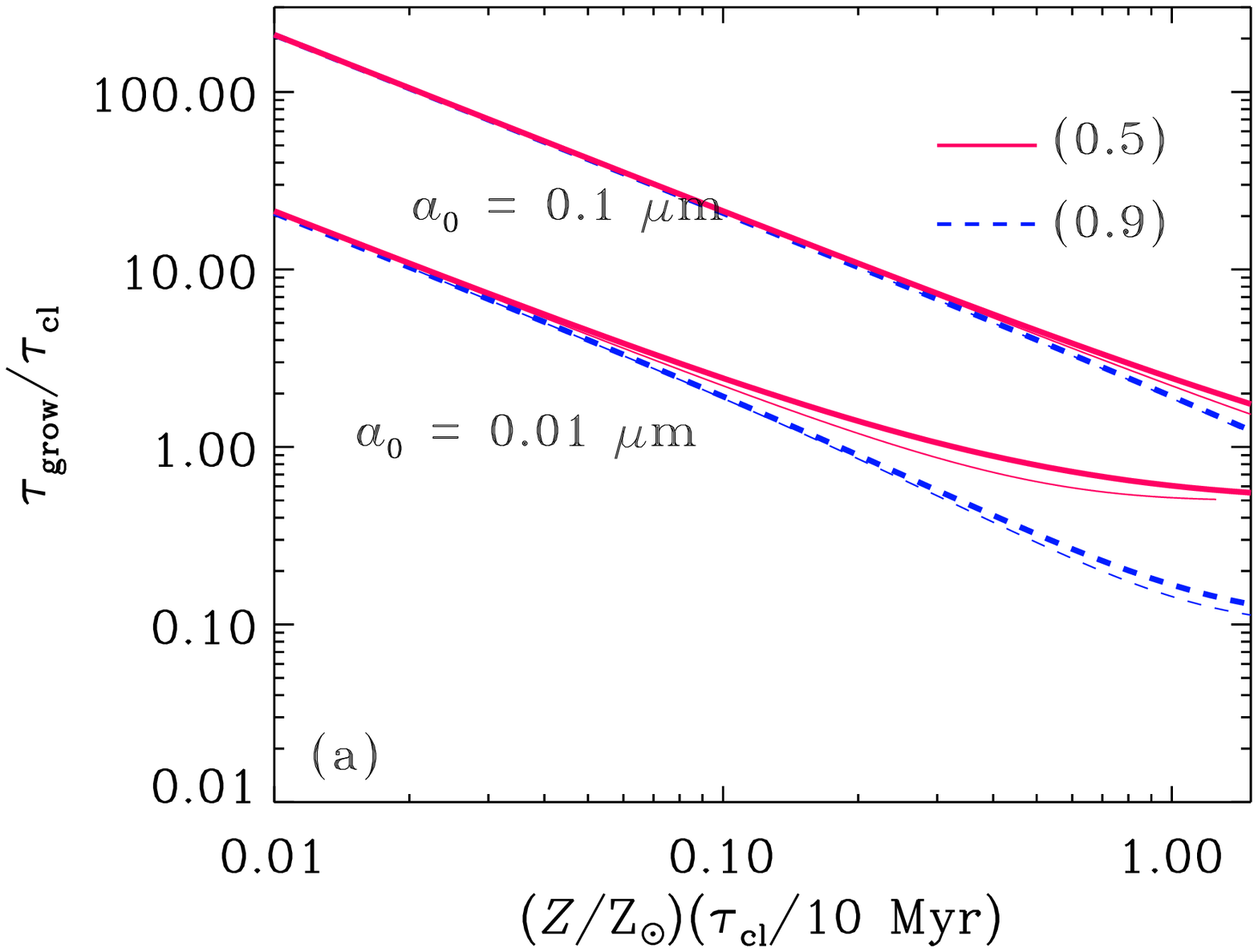}
\includegraphics[width=0.45\textwidth]{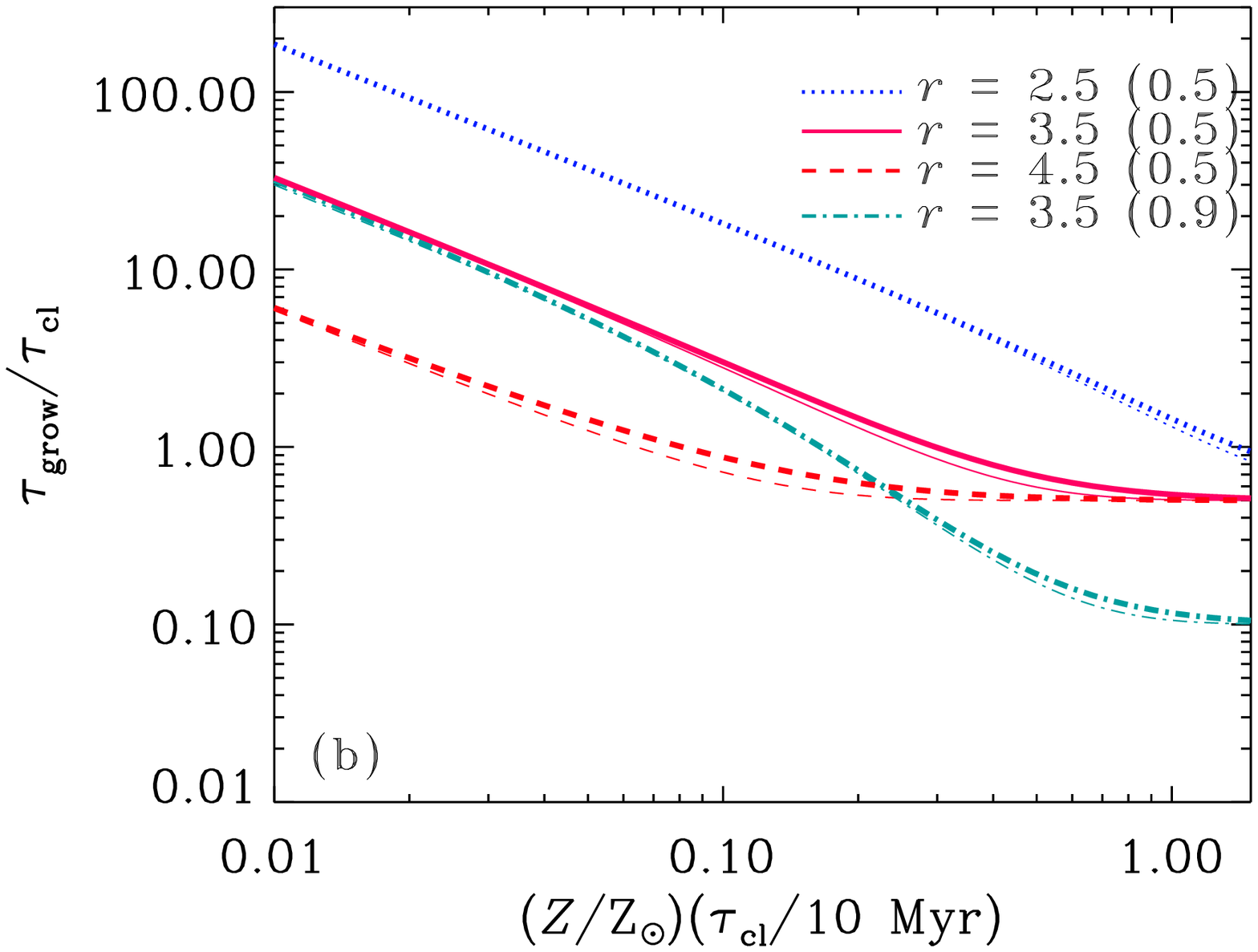}
 \caption{Same as Fig.\ \ref{fig:taugrow} but we show the
 approximate results with equation (\ref{eq:approx}).
 The thick lines show the approximate relations
 and, while the
 thin lines are calculated by the exact formulation
 used to produce the results in Fig.\ \ref{fig:taugrow}.
 }
 \label{fig:taugrow_fit}
\end{figure}

In Fig.\ \ref{fig:taugrow}b, we overlay the relation
predicted by equation (\ref{eq:taugrow_smallZ}) for
$r=3.5$ (the upper thin dotted line, which should be
compared with the dot-dashed line for the exact
solution). Although it gives a good approximation for
low metallicities, the discrepancy becomes
significant at
$(Z/Z_{\sun})(\tau_\mathrm{cl}/10~\mathrm{Myr})\ga 0.1$.
This is because the increase of
dust surface area by
accretion further accelerates the accretion of
gas phase metals.
Such an effect of acceleration of grain growth is
not taken into account in deriving
equation (\ref{eq:taugrow_smallZ}), but is
included in the second and
the third order terms of ${A}$ in
equation (\ref{eq:taugrow_taucl}), where $A$
can be approximated
by equation (\ref{eq:A_lowZ}) if the saturation of
grain growth by the depletion of gas-phase metals
is neglected.
Thus, we expect that the approximation
of $\tau_\mathrm{grow}$ becomes better
than equation (\ref{eq:taugrow_smallZ})
if we combine equations (\ref{eq:taugrow_taucl})
and (\ref{eq:A_lowZ}).
The result predicted by this combination
is also shown in
Fig.~\ref{fig:taugrow}b (thin dotted line).
It fits the
exact solution better than
equation (\ref{eq:taugrow_smallZ}), but it underestimates
$\tau_\mathrm{grow}$ for large metallicities.
This underestimate comes from the fact that we neglected the
depletion of metals by the grain growth. This depletion
effect occurs if the cloud lifetime is comparable or larger
than the typical grain growth time-scale. Thus, we propose
the following approximate formula to include the depletion
effect:
\begin{eqnarray}
\frac{\tau_\mathrm{grow}}{\tau_\mathrm{cl}}\simeq
\frac{\xi\,\langle{a}^3\rangle_0}
{3y\langle{a}^2\rangle_0+3y^2\langle{a}\rangle_0
+y^3}+(1-\xi ),\label{eq:approx}
\end{eqnarray}
where $y\equiv a_0\xi\tau_\mathrm{cl}/\tau$
(Equation \ref{eq:A_lowZ}).
In Fig.\ \ref{fig:taugrow_fit}, we show this approximate
formula in comparison with the exact results. We
observe that the approximation is fairly good.
By using equation (\ref{eq:taugrow_taucl}),
$\beta =\xi (\tau_\mathrm{grow}/\tau_\mathrm{cl})^{-1}$.
Thus, we adopt the following approximation derived from
equation (\ref{eq:approx}):
\begin{eqnarray}
\beta\simeq
\left[\frac{\langle{a}^3\rangle_0}
{3y\langle{a}^2\rangle_0+3y^2\langle{a}
\rangle_0+y^3}+\frac{1-\xi}{\xi}
\right]^{-1}\label{eq:beta_approx}
\end{eqnarray}
with $y\equiv a_0\xi\tau_\mathrm{cl}/\tau$. This
approximate formula has a merit that we do not need
to solve the differential equation for the depletion of
gas-phase metals (equation \ref{eq:dndt}).
Therefore, we hereafter use equations (\ref{eq:approx})
and (\ref{eq:beta_approx}) to estimate $\tau_\mathrm{grow}$
and $\beta$, respectively.

\subsection{Relation to the star formation}
\label{subsec:dMdt2}

Galaxies are enriched by metals and dust as a result
of star formation activities. Thus, it is usually
convenient to model the quantities in terms of
the star formation. In fact, dust growth by accretion
should be strongly related to star formation, since both
occur predominantly in dense clouds in galaxies
\citep{dwek98}. Here we make use of $\beta$ and
connect the star formation rate with the grain growth rate
by accretion.

If the fraction $\epsilon$ of the cloud mass is
converted into stars after the cloud lifetime
$\tau_\mathrm{cl}$, the
star formation rate, $\psi$, is written as
\begin{eqnarray}
\psi =
\frac{\epsilon X_\mathrm{cl}M_\mathrm{gas}}
{\tau_\mathrm{cl}},\label{eq:sfr}
\end{eqnarray}
where $M_\mathrm{gas}$ is the total gas mass in
the galaxy. Then, equation (\ref{eq:dMdt2}) can be
written by using the star formation rate as
\begin{eqnarray}
\left[\frac{\mathrm{d}M_\mathrm{dust}}{\mathrm{d}t}
\right]_\mathrm{acc}=
\frac{\beta\mathcal{D}\psi}{\epsilon},\label{eq:dMdt3}
\end{eqnarray}
where
$\mathcal{D}\equiv M_\mathrm{dust}/M_\mathrm{gas}$
is the dust-to-gas ratio.

\subsection{Recipe for implementation of grain growth}
\label{subsec:recipe}

Here we summarize the recipe to include the grain
growth by accretion into dust enrichment models.
There are two methods as described in
sections \ref{subsec:dMdt1} and \ref{subsec:dMdt2},
called Method I and Method II, respectively.

\begin{description}
\item[\textbf{Method I}] Assume $X_\mathrm{cl}$
(the mass fraction of clouds hosting grain growth)
and $\tau_\mathrm{cl}$ (the lifetime of these
clouds). Calculate also $\langle{a}^\ell\rangle$
for $\ell =1$, 2, and 3 for a given grain size
distribution (or use the values in
Table \ref{tab:model}). Then, under any metallicity,
$\tau$ (the typical grain growth time-scale) is
calculated by equation (\ref{eq:tau_sil}) for
silicate, by equation (\ref{eq:tau_gra}) for graphite
or by equation (\ref{eq:tau}) in general, and
$y=a_0\xi\tau_\mathrm{cl}/\tau$ is also obtained.
Note that $\xi$ (the fraction of metals in gas phase)
should be calculated within the framework of dust
enrichment (Section \ref{subsec:enrichment}).
Finally use equation (\ref{eq:taugrow_taucl})
(more conveniently equation \ref{eq:approx} for
an approximate formula) to obtain
$\tau_\mathrm{grow}$, which should be used in
equation (\ref{eq:dMdt1}) to obtain
$[\mathrm{d}M_\mathrm{dust}/\mathrm{d}t
]_\mathrm{acc}$.
\item[\textbf{Method II}] Assume
$\tau_\mathrm{cl}$ (the lifetime of clouds) and
$\epsilon$ (the star formation efficiency in these
clouds). Calculate also $\langle{a}^\ell\rangle$
for $\ell =1$, 2, and 3 for a given grain size
distribution (or use the values in
Table \ref{tab:model}). Then, under any metallicity,
$\tau$ is
calculated by equation (\ref{eq:tau_sil}) for
silicate or equation (\ref{eq:tau_gra}) for graphite,
and $y=a_0\xi\tau_\mathrm{cl}/\tau$ is also obtained.
Note that $\xi$
should be calculated by a chemical enrichment
model. Finally use equation (\ref{eq:beta})
(or more conveniently equation \ref{eq:beta_approx}
for an approximate formula) to obtain
$\beta$, which should be used in
equation (\ref{eq:dMdt3}).
\end{description}

\subsection{Dust enrichment model}
\label{subsec:enrichment}

We demonstrate that our results so far are really
implemented into galaxy evolution models with
dust enrichment. The aim here is not to construct
an elaborate chemical evolution model but to focus
on the effect of grain growth. We adopt a one-zone
closed-box model of dust/metal enrichment by
\citet{hirashita00b}
\citep[see also][]{dwek98,lisenfeld98}. The model
treats the evolutions of total gas, metals, and
dust masses ($M_\mathrm{gas}$, $M_{Z}$, and
$M_\mathrm{dust}$, respectively) in the galaxy.
In this model, the metals include not only gas
phase elements but also dust. The equations are
written as
\begin{eqnarray}
\frac{\mathrm{d}M_\mathrm{gas}}{\mathrm{d}t} & = &
-\psi +E,\label{eq:dMgdt}\\
\frac{\mathrm{d}M_{Z}}{\mathrm{d}t} & = &
-Z\psi+E_{Z}\\
\frac{\mathrm{d}M_\mathrm{dust}}{\mathrm{d}t} & = &
-\mathcal{D}\psi +f_\mathrm{in}E_{Z}-
\frac{M_\mathrm{dust}}{\tau_\mathrm{SN}}+
\left[\frac{\mathrm{d}M_\mathrm{dust}}{\mathrm{d}t}
\right]_\mathrm{acc},\label{eq:dMddt}
\end{eqnarray}
where $E$ and $E_Z$ are the rate of the total
injection of mass (gas + dust) and metal mass
from stars, respectively, $f_\mathrm{in}$
is the dust condensation efficiency of the metals in
the stellar ejecta, and $\tau_\mathrm{SN}$ is the
time-scale of dust destruction by SN shocks.
For simplicity, we do not treat individual element X
but treat the entire metals; however, if we are
interested in a specific dust-composing element X,
we can replace the relevant quantities with ones
specific for element X. The simple
treatments above are sufficient to
demonstrate that our scheme
of grain growth in clouds for any grain size distribution
is really applicable to the galaxy evolution models.
We refer to other papers
\citep[e.g.][]{dwek98,zhukovska08,gall11} for more
detailed dust enrichment models.

Since we are not interested in the detailed history
of dust production in stellar ejecta, we adopt the
instantaneous recycling
approximation; that is, a star with $m>m_t$
($m$ is the zero-age stellar mass, and $m_t$ is
the turn-off mass at age $t$) dies
instantaneously after its birth, leaving a remnant of
mass $w_m$. Once the initial mass function (IMF) is
fixed, the returned fraction of the mass
from formed stars, $\mathcal{R}$, and the
mass fraction of metals that is newly produced
and ejected by stars, $\mathcal{Y}_{Z}$, are evaluated.
Using these quantities, we write
\begin{eqnarray}
E & = & \mathcal{R}\psi ,\\
E_{Z} & = & (\mathcal{R}Z+
\mathcal{Y}_{Z})\psi .
\end{eqnarray}
We adopt $\mathcal{R}=0.18$ and
$\mathcal{Y}_\mathrm{X}=0.013$
(Appendix).
For $f_\mathrm{in}$, we examine two cases:
$f_\mathrm{in}=0.1$ and 0.01, which correspond
to the fiducial and the lower efficiency cases
in \citet{inoue11}, respectively.

For the time-scale of dust destruction by SNe, we
adopt an expression
$\tau_\mathrm{SN}=M_\mathrm{gas}/(\epsilon_\mathrm{s}
M_\mathrm{s}\gamma )$, where $\epsilon_\mathrm{s}$
and $M_\mathrm{s}$ are the dust destruction efficiency
and the gas mass swept by a single high-velocity SN
blast, respectively, and $\gamma$ is the SN rate
\citep{mckee89}. Since we are interested in objects
whose time-scale of star formation is much longer than
$10^7$ yr, it is assumed that the SN rate is
proportional to the star formation rate
(equation \ref{eq:snr}). We adopt
$\epsilon_\mathrm{s}M_\mathrm{s}=1300$ M$_{\sun}$
\citep{mckee89}. Then we obtain
\begin{eqnarray}
\tau_\mathrm{SN}=
\frac{M_\mathrm{gas}}{\beta_\mathrm{SN}\psi},
\label{eq:tauSN}
\end{eqnarray}
where
$\beta_\mathrm{SN}\equiv\epsilon_\mathrm{s}M_\mathrm{s}
\gamma /\psi\simeq 9.65$.
{As pointed out by \citet{jones11}, $\epsilon$
is uncertain by a factor of $\sim 2$ and is dependent
on the assumed grain composition
\citep[see also][]{serra08}.}

Equations (\ref{eq:dMgdt})--(\ref{eq:dMddt}) are
converted to the time evolution of the metallicity
$Z=M_\mathrm{Z}/M_\mathrm{gas}$ and the dust-to-gas
ratio $\mathcal{D}=M_\mathrm{dust}/M_\mathrm{gas}$
as
\begin{eqnarray}
\frac{M_\mathrm{gas}}{\psi}\,
\frac{\mathrm{d}Z}{\mathrm{d}t} & \hspace{-3mm}= &
\hspace{-3mm}\mathcal{Y}_Z,\\
\frac{M_\mathrm{gas}}{\psi}\,
\frac{\mathrm{d}\mathcal{D}}{\mathrm{d}t} &
\hspace{-3mm}= & \hspace{-3mm}
f_\mathrm{in}(\mathcal{R}Z+\mathcal{Y}_Z)-
(\beta_\mathrm{SN}+\mathcal{R})\mathcal{D}\nonumber\\
& & +\frac{1}{\psi}\left[
\frac{\mathrm{d}M_\mathrm{dust}}{\mathrm{d}t}
\right]_\mathrm{acc},
\end{eqnarray}
where we should evaluate
$[\mathrm{d}M_\mathrm{dust}/\mathrm{d}t]_\mathrm{acc}$
according to Method I or II in
Section \ref{subsec:recipe}.
It is convenient to combine the above two equations
to obtain the relation between $\mathcal{D}$ and $Z$:
\begin{eqnarray}
\mathcal{Y}_\mathrm{Z}
\frac{\mathrm{d}\mathcal{D}}{\mathrm{d}Z} & = &
f_\mathrm{in}(\mathcal{R}Z+\mathcal{Y}_Z)-
(\beta_\mathrm{SN}+\mathcal{R})\mathcal{D}\nonumber\\
& &
+\frac{1}{\psi}\left[
\frac{\mathrm{d}M_\mathrm{dust}}{\mathrm{d}t}
\right]_\mathrm{acc}.\label{eq:dDdZ}
\end{eqnarray}

In Method I,
$[\mathrm{d}M_\mathrm{dust}/\mathrm{d}t]_\mathrm{acc}/
\psi=\mathcal{D}\xi (\tau_\mathrm{SF}/
\tau_\mathrm{grow})$ (equation~\ref{eq:dMdt1}), where
$\tau_\mathrm{SF}\equiv X_\mathrm{cl}M_\mathrm{gas}
/\psi$
is the star formation time-scale. In Method II,
$[\mathrm{d}M_\mathrm{dust}/\mathrm{d}t]_\mathrm{acc}/
\psi=\beta\mathcal{D}/\epsilon$
(equation \ref{eq:dMdt3}). A large $\tau_\mathrm{SF}$
in Method I is equivalent with a small $\epsilon$ in
Methods II; that is, a small
star formation efficiency means a long star formation
time-scale. Because of this simple equivalence, we
hereafter concentrate on Method~II.
Adopting Method II (i.e.\ equation \ref{eq:dMdt3}),
equation (\ref{eq:dDdZ}) is reduced to
\begin{eqnarray}
\mathcal{Y}_\mathrm{Z}
\frac{\mathrm{d}\mathcal{D}}{\mathrm{d}Z} & = &
f_\mathrm{in}(\mathcal{R}Z+\mathcal{Y}_Z)-
(\beta_\mathrm{SN}+\mathcal{R}-\beta /\epsilon )
\mathcal{D}.\label{eq:dDdZ_final}
\end{eqnarray}

\citet{lada10} show that the star formation
efficiency in molecular clouds is roughly
0.1. They also mention that molecular clouds survive after
the star formation activity
over the last 2 Myr. The comparison with the
age of stellar clusters associated with molecular
clouds indicates that the lifetime of clouds is
$\sim 10$ Myr \citep{leisawitz89,fukui10}.
Thus, we hereafter assume
$\epsilon =0.1$ and $\tau_\mathrm{mol}=10$ Myr
as standard values. For the initial condition, we
assume $\mathcal{D}=0$ and $Z=0$.

\begin{figure}
\includegraphics[width=0.45\textwidth]{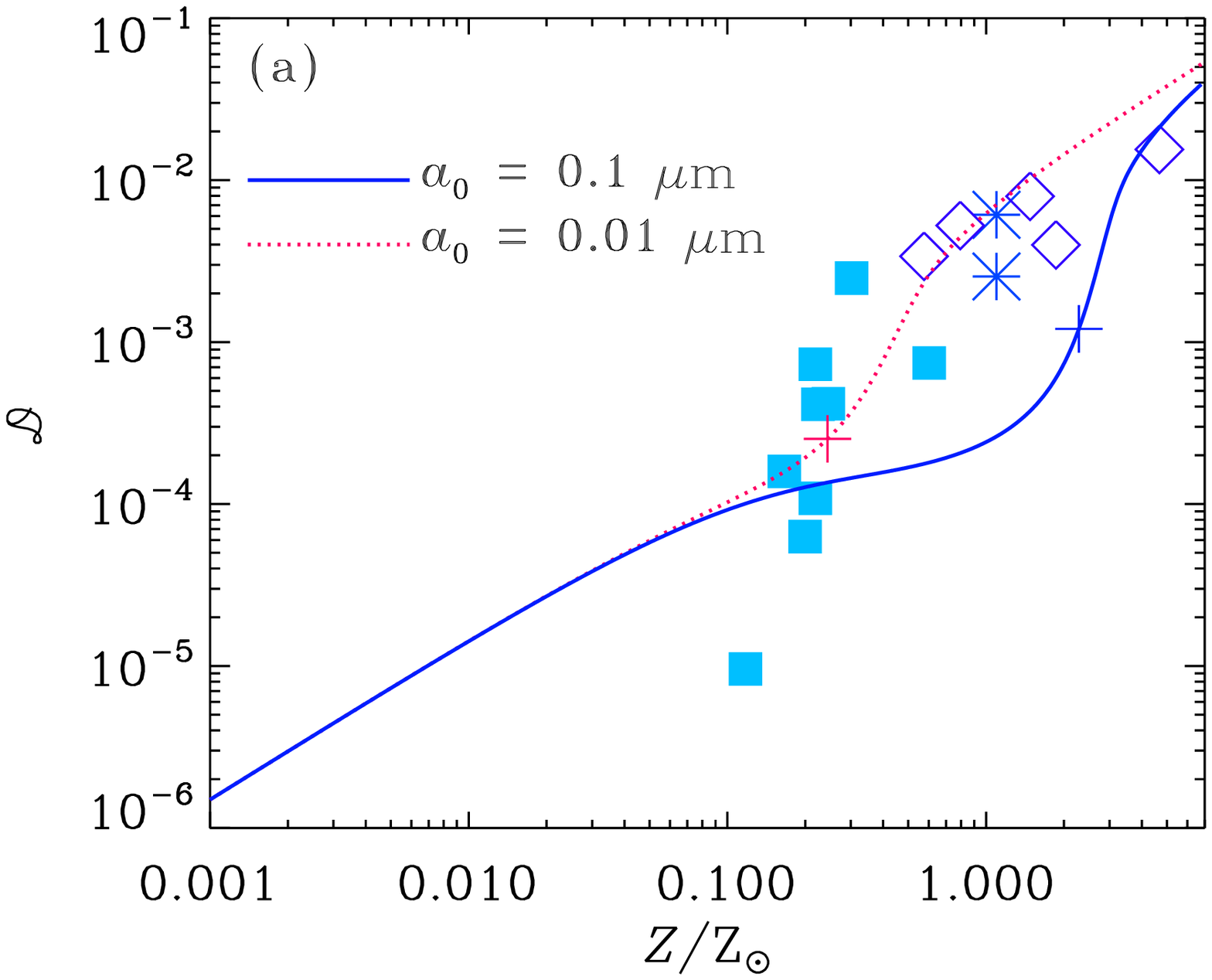}
\includegraphics[width=0.45\textwidth]{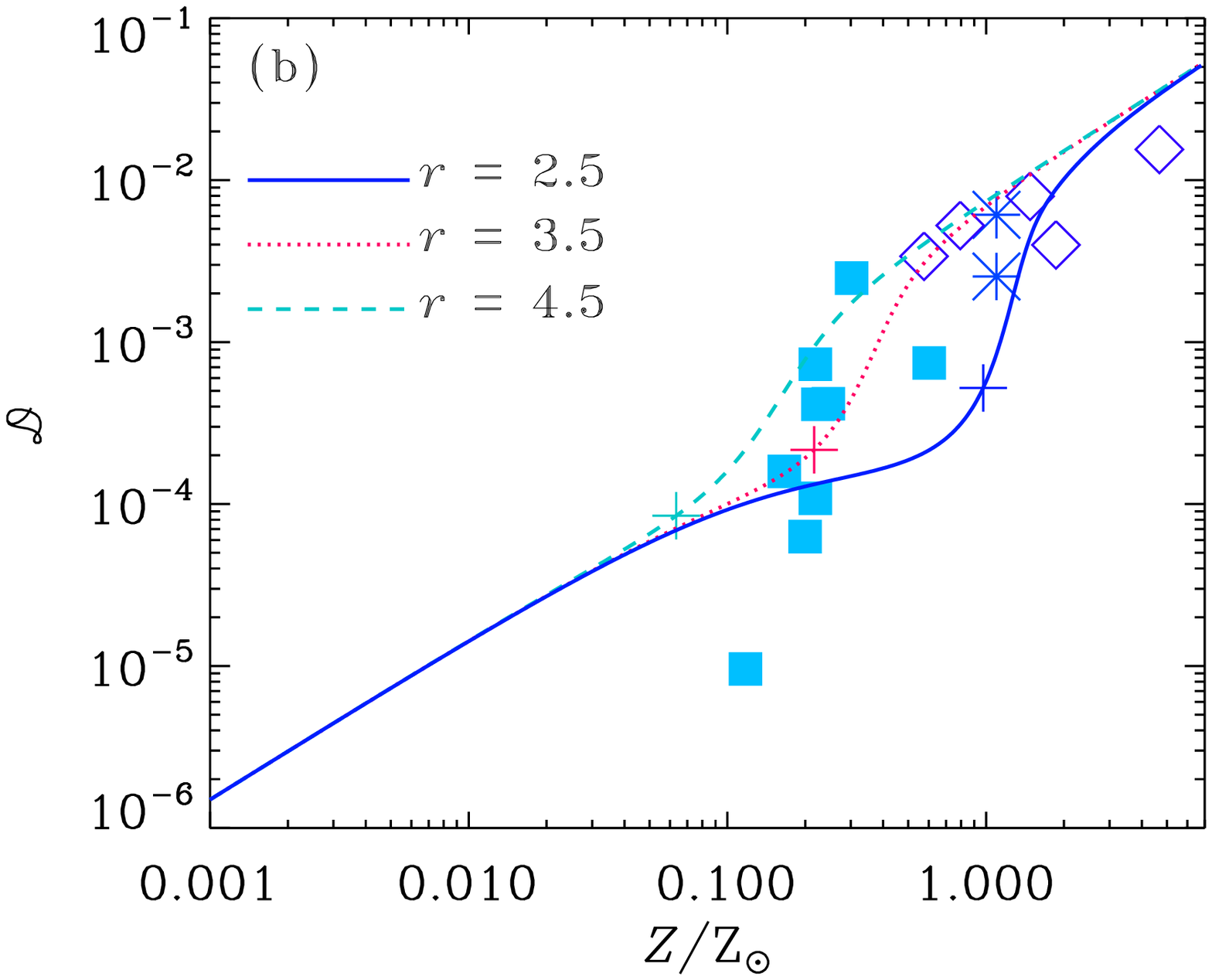}
 \caption{Relation between dust-to-gas ratio
$\mathcal{D}$ and metallicity $Z$ for various grain
size distributions with $f_\mathrm{in}=0.1$.
(a) $\delta$ function grain size distributions (the solid
and dotted lines present the results for $a_0=0.1$ and
0.01\,$\micron$, respectively).
(b) Power-law grain size distributions (the solid, dotted,
and dashed lines show the results for
$r=2.5$, 3.5, and 4.5, respectively).
The filled squares and asterisks
represent the observational data for dwarf and
spiral galaxies observed by \textit{AKARI},
respectively \citep{hirashita09b,suzuki07,sun11}.
The open diamonds are taken from
\citet{issa90} as a spiral galaxy sample.
The cross on each line marks the critical metallicity.
 }
 \label{fig:dg_metal}
\end{figure}

\begin{figure}
\includegraphics[width=0.45\textwidth]{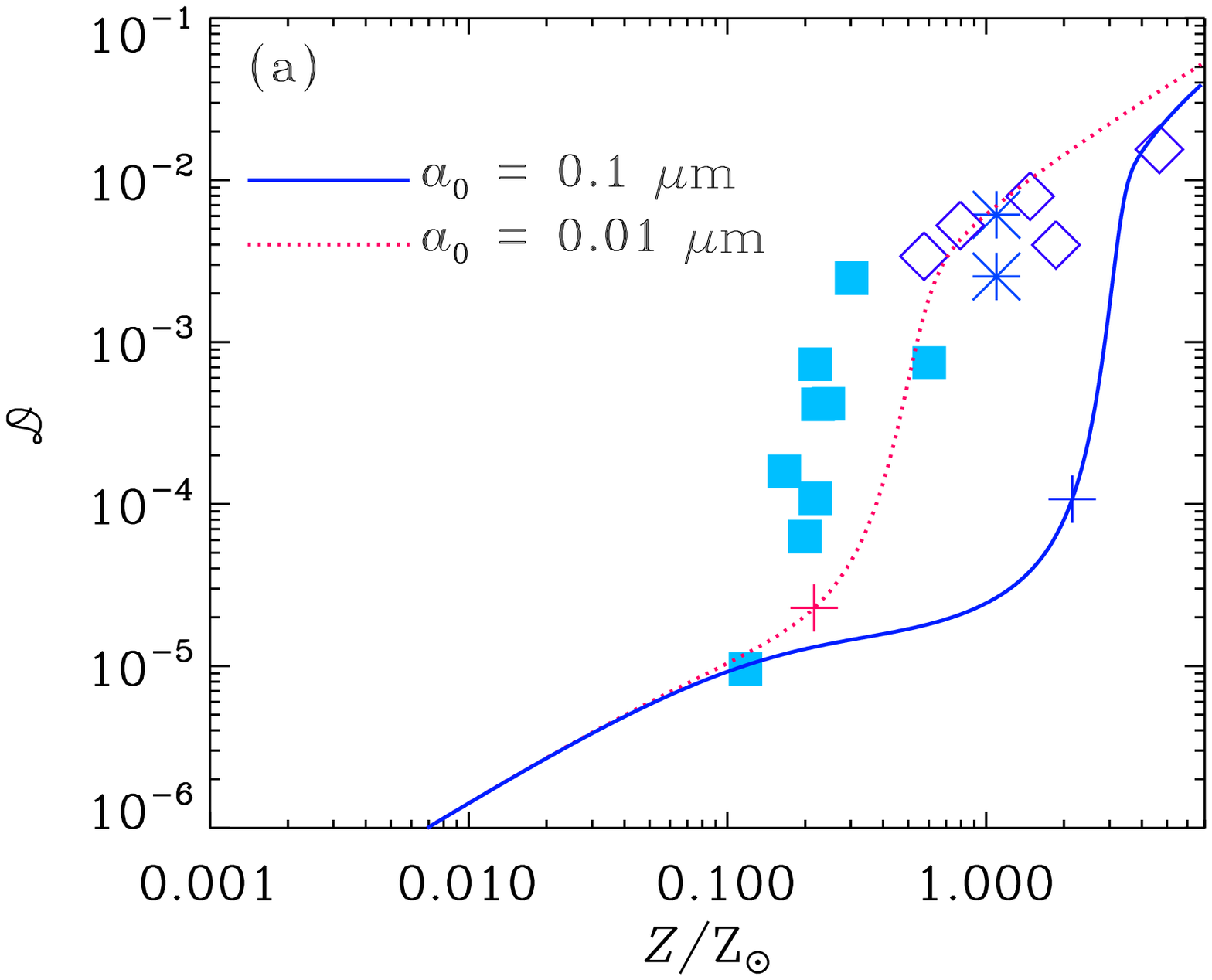}
\includegraphics[width=0.45\textwidth]{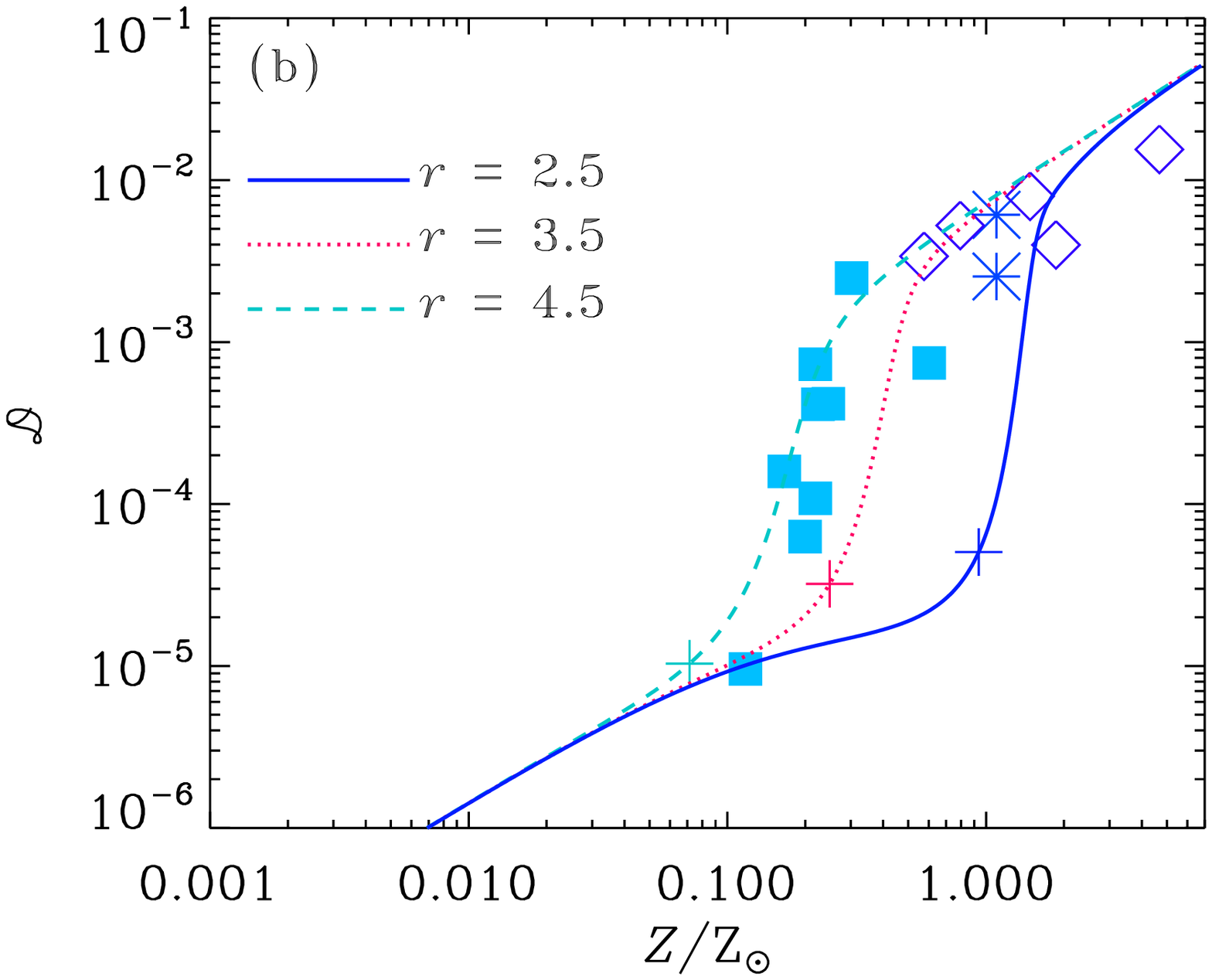}
 \caption{Same as Fig.\ \ref{fig:dg_metal} but for
 $f_\mathrm{in}=0.01$.
 }
 \label{fig:dg_metal_fin0.01}
\end{figure}

The relations between dust-to-gas ratio and
metallicity are shown in Fig.\ \ref{fig:dg_metal}
for $f_\mathrm{in}=0.1$ and in
Fig.\ \ref{fig:dg_metal_fin0.01} for
$f_\mathrm{in}=0.01$.
{We assume Z$_{\sun}=0.015$
\citep{lodders03}.}
At low metallicity levels, the solution of
equation (\ref{eq:dDdZ_final}) is approximated by
$\mathcal{D}\sim f_\mathrm{in}Z$. This is why
the dust-to-gas ratio in the low-metallicity regime
is lower in Fig.\ \ref{fig:dg_metal_fin0.01} than in
Fig.\ \ref{fig:dg_metal}. Above a certain metallicity
level, a rapid increase of dust-to-gas ratio
occurs because of
the grain growth in
clouds. The metallicity level
at which this growth occurs is very sensitive
to the grain size distribution.

The observational data of nearby galaxies are
also shown for comparison. For the uniformity of
data, we select
the samples observed by \textit{AKARI}:
blue compact dwarf galaxies in
\citet{hirashita09b} and spiral galaxies
(M\,81 from \citealt{sun11} and
M\,101 from \citealt{suzuki07}).
The dust masses are estimated from
90 $\micron$ and 140 $\micron$ data
by using the mass absorption coefficient
in \citet{hirashita09b}. For the gas mass of
the dwarf galaxies and M\,81,
we adopt the H \textsc{i} mass, since the
molecular mass
is negligible or not detected.
The H \textsc{i} masses of M\,81 and of
the dwarf galaxies are taken from
\citet{walter08} and
in \citet{hirashita09b}, respectively. For M\,101,
we adopt the sum of neutral and molecular
gas masses compiled in \citet{suzuki07}.
The oxygen abundance is adopted for the
indicator of the metallicity, and the solar
abundance is assumed to be
{$12+\log\mathrm{(O/H)}=8.69$
\citep{lodders03}}. The
oxygen abundances of the dwarf galaxies are
compiled in \citet{hirashita09b}, and those of
the spiral galaxies at the half-light radius
are taken from
\citet{garnett02}. Typical errors of the
observational quantities are comparable to
the size of symbols in the figures.
To ensure that the \textit{AKARI} results
are not systematically different from other
results, we add the data in \citet*{issa90},
who estimated the dust mass from the extinction.
The overall trend of the
data is reproduced by the models.

\citet{asano11} also include a sample whose dust
mass is derived from \textit{Spitzer} observations
by \citet{engelbracht08} and show that the relation
between dust-to-gas ratio and metallicity does not
significantly change. The inclusion of \textit{Herschel}
data may boost the dust abundance because
of the possible contribution from very cold dust
especially in dwarf galaxies \citep{grossi10}.
\citet{galliano05} also find a large contribution
from very cold dust in the submillimetre for
some dwarf galaxies. However, since
the modeling of submillimetre emission may be
significantly affected by the assumed emissivity
index of large grains, we do not use the submillimetre
data in this paper. We should keep in mind that
that inclusion of submillimetre data can rather
\textit{decrease} the dust mass especially for
dust-rich galaxies because the
dust temperature estimate becomes better
\citep{galametz11}. In their models, the
\textit{Spitzer} 160 $\micron$ data are used;
thus, it is not still unclear if there is a
discrepancy between the dust temperatures estimated
by \textit{AKARI} and those estimated by including
submillimetre data.
The dust mass in M\,81 derived by the \textit{Herschel}
observation ($3.4\times 10^7$ M$_\odot$;
\citealt{bendo10}) is similar to that estimated by
the \textit{AKARI} observation adopted in
this paper ($3.2\times 10^7$ M$_\odot$; \citealt{sun11}).

\section{Discussion}\label{sec:discussion}

\subsection{Effects of grain size distribution}

We have shown that the grain size distribution
significantly affects the evolution of dust mass
above a certain metallicity level where the grain
growth in clouds is activated. In fact, as shown
in Figs.~\ref{fig:dg_metal} and
\ref{fig:dg_metal_fin0.01}, the difference in the
grain size distribution makes a significant
imprint in the relation between dust-to-gas ratio
and metallicity. This comes from the dependence
of $\beta$ on metallicity. As shown in
equation (\ref{eq:taugrow_taucl}),
$\beta\propto(\tau_\mathrm{grow}
/\tau_\mathrm{cl})^{-1}$.
Fig.\ \ref{fig:taugrow} shows that
$\tau_\mathrm{grow}/\tau_\mathrm{cl}$ is a
decreasing function of metallicity. Thus,
$\beta$ increases
as the system is enriched with metals, and if
the last term in equation (\ref{eq:dDdZ_final})
becomes positive at a certain metallicity, the dust
mass grows in a nonlinear way. Because $\beta$
(or $\tau_\mathrm{grow}/\tau_\mathrm{cl}$) is
sensitive to the grain size distribution, the
resulting relation between dust-to-gas ratio and
metallicity depends largely on the grin size
distribution. In other words, the metallicity level
at which the
grain growth in clouds is activated
varies sensitively with the grain size distribution.
This metallicity
level is called `critical metallicity', which is
quantified and discussed in the next subsection.

\subsection{Critical metallicity for the grain growth}
\label{subsec:Zcr}

\citet{asano11} and \citet{inoue11} have shown
that the grain growth by accretion dominates the
grain abundance if the metallicity is larger than
a certain `critical metallicity'.
As mentioned in the previous subsection,
the nonlinear increase of
dust-to-gas ratio by grain growth is realized when
$-\beta_\mathrm{SN}-\mathcal{R}+\beta /\epsilon >0$
in equation (\ref{eq:dDdZ_final}). Since
$\beta_\mathrm{SN}\gg\mathcal{R}$, the critical
metallicity, $Z_\mathrm{cr}$, can be estimated by
the metallicity at which $\beta$ as a function of
metallicity realizes
\begin{eqnarray}
\beta (Z_\mathrm{cr})=\epsilon\beta_\mathrm{SN},
\label{eq:beta_cr}
\end{eqnarray}
where $\beta (Z)$ indicates that $\beta$ is
a function of metallicity. In our models,
$\epsilon =0.1$ and
$\beta_\mathrm{SN}=9.65$, i.e.,
$\beta (Z_\mathrm{cr})=0.965$. Therefore, if the
metallicity is so high that the grain growth in
individual clouds
increases the dust mass by 0.965 times the original
dust amount, the grain growth becomes prominent.
The critical metallicity varies by a factor of
$\sim 1.5$
if we change $\beta_\mathrm{SN}$ by
a factor of 2 (i.e.\ uncertainty in $\beta_\mathrm{SN}$
from the assumed species; \citealt{jones11}).

The position for the critical metallicity is marked on
each line in Figs.\ \ref{fig:dg_metal} and
\ref{fig:dg_metal_fin0.01} and listed in
Table \ref{tab:Zcr} (for $f_\mathrm{in}=0.1$
but $Z_\mathrm{cr}$ only decrease by $\sim 10$--20 per cent
for $f_\mathrm{in}=0.01$). Indeed the dust-to-gas ratio
rapidly increases if the metallicity becomes
larger than the critical metallicity.
This supports \citet{inoue11}'s view that the
activation of grain growth is driven by the
metallicity. In this paper we have found that
the critical metallicity is sensitive to the grain
size distribution. If the major part of the grains
are as large as $\sim 0.1~\micron$, the grain
growth is activated above $\sim 2$ Z$_{\sun}$, which
is too large to explain the large grain abundance
in the objects with sub-solar metallicities.
If the typical grain size is as small as
$\sim 0.01~\micron$ or the grain size distribution
is described by a power law with $r\ga 3.5$, the
large grain abundance in sub-solar metallicity
galaxies can be naturally explained by the
efficient grain growth in clouds.
Therefore, the evolutionary history of grain
size distribution is important to understand
the grain abundance in galaxies.

\begin{table}
\centering
\begin{minipage}{60mm}
\caption{Critical metallicities for grain growth.}
\label{tab:Zcr}
    \begin{tabular}{cccccc}
     \hline
     ${n}\,^\mathrm{a}$ & $a_0$ & $r$ & $a_\mathrm{min}$
     & $a_\mathrm{max}$ &
     $Z_\mathrm{cr}$\\
      & [$\micron$] & & [$\micron$] & [$\micron$] & [Z$_{\sun}$]\\
     \hline 
     $\delta$ & 0.1 & --- & ---
       & --- & 2.3 \\
     $\delta$ & 0.01 & --- & ---
       & --- & 0.24\\
     p & 0.1 & 3.5 & 0.001 & 0.25 & 0.22\\
     p & 0.1 & 2.5 & 0.001 & 0.25 & 0.97\\
     p & 0.1 & 4.5 & 0.001 & 0.25 & 0.063\\
     p & 0.1 & 3.5 & 0.0003 & 0.25 & 0.094\\
     p & 0.1 & 3.5 & 0.003 & 0.25 & 0.45\\
     \hline
    \end{tabular}

\medskip

$^\mathrm{a}$ Grain size distribution: ``p'' for
the power law and ``$\delta$'' for the $\delta$ function.
\end{minipage}
\end{table}

{
By an analysis of infrared dust emission spectra,
\citet{galliano05} show that the grain size
distributions of some metal-poor dwarf galaxies are
biased toward
smaller grains compared with the Galactic case
(i.e.\ $r\simeq 3.5$). Therefore, the critical metallicity
for these dwarf galaxies is expected to be lower than
their metallicities: in other words, the grain growth
in clouds is expected to contribute significantly to
the dust abundance in these galaxies.
It is interesting to point out that $r=4.5$ explains
the data points of metal-poor galaxies better than
$r=3.5$. This is consistent with \citet{galliano05}'s
conclusion that the grain size distribution is biased
to small sizes.
}

After the grain growth, all the lines in
Figs.\ \ref{fig:dg_metal} and
\ref{fig:dg_metal_fin0.01}
finally converge. This is explained as follows.
If the metallicity is high enough, the grain growth
is regulated by the lifetime of clouds
and is independent of grain size distribution.
In other words, $\beta\sim\xi/(1-\xi )$ if the
metallicity becomes high enough
(i.e.\ $y$ is large enough in
equation \ref{eq:beta_approx}).
Using $\xi =1-\mathcal{D}/Z$ and
equation (\ref{eq:dDdZ_final}), we obtain the
following estimate in the case where the grain
growth dominates the increase of the dust content:
\begin{eqnarray}
\mathcal{Y}_Z\frac{\mathrm{d}\xi}{\mathrm{d}Z}
& \sim & (\beta_\mathrm{SN}-\beta /\epsilon )
\frac{\mathcal{D}}{Z^2}\nonumber\\
& \sim & \left(\beta_\mathrm{SN}
-\frac{1}{\epsilon}\,\frac{\xi}{1-\xi}\right)
\frac{\mathcal{D}}{Z^2},
\end{eqnarray}
which is valid for $Z>Z_\mathrm{c}$
(we only adopt the dominant terms). Therefore,
if $\xi$ is so small (large) that the right-hand
side of this equation is positive (negative),
$\xi$ tends to
increase (decrease) as $Z$ increases. This
means that $\xi$ tends to approach the value
that makes the right-hand side to be zero;
that is,
\begin{eqnarray}
\xi\sim
\frac{\epsilon\beta_\mathrm{SN}}
{1+\epsilon\beta_\mathrm{SN}}~~~\mbox{for}~
Z\gg Z_\mathrm{cr}.
\end{eqnarray}
For the values adopted in our models
($\epsilon =0.1$ and $\beta_\mathrm{SN}=9.65$),
$\xi\sim 0.49$.
This means that the fraction of metals in
dust phase is about 0.5 at metallicities
much higher than the critical metallicity.
{If we consider the uncertainty in
$\beta_\mathrm{SN}$ by a factor of 2
because of material difference
\citep{jones11},
$\xi\sim 0.33$--0.66.}

It is interesting that $\xi$ at $Z\gg Z_\mathrm{cr}$
only depends on $\epsilon\beta_\mathrm{SN}$.
Using the definition of
$\beta_\mathrm{SN}=
\epsilon_\mathrm{s}M_\mathrm{s}\gamma /\psi$
(below equation \ref{eq:tauSN}) and
equation (\ref{eq:sfr}), we obtain
\begin{eqnarray}
\epsilon\beta_\mathrm{SN}=
\frac{\epsilon_\mathrm{s}M_\mathrm{s}\gamma}
{X_\mathrm{cl}M_\mathrm{gas}/\tau_\mathrm{cl}}.
\end{eqnarray}
This is the ratio between the gas mass swept by
SN shocks per unit time multiplied by the efficiency of
dust destruction in SN shocks, and the formation rate
of clouds. In other words, this is the ratio between the
dust destruction rate by SN shocks and the
grain growth rate in clouds. Therefore, it is
natural that the final fraction of metals in dust
phase can be described by the balance between
the dust destruction by SN shocks and the dust formation
in clouds.

\citet{inoue11} also obtained a similar expression
for the critical metallicity. We have
confirmed his conclusion that the dust-to-metal
ratio approaches a constant value: this behaviour
is called self-regulation by \citet{inoue11}.
The difference between our formulation and
his is that he treated $\tau_\mathrm{SF}$
as a parameter independent of $\tau_\mathrm{grow}$
while we connect these two parameters through
the star formation efficiency $\epsilon$.
Since both star formation and grain growth
occur in molecular clouds, these two processes
are not independent. Moreover, if the grain growth
is efficient enough, the dust
growth time-scale is limited by the lifetime
of clouds, which is independent of the
grain size distribution. Therefore, the
dust-to-metal ratio does not depend on the grain size
distribution for $Z\gg Z_\mathrm{cr}$, and it
depends only on
$\epsilon\beta_\mathrm{SN}$.

\subsection{Significance in galaxy evolution}

The first dust should be produced by SNe in the
death of massive stars. Because of shock
destruction in SNe, the dust sizes may be biased
to $a\ga 0.1~\micron$ \citep{bianchi07,nozawa07}.
\citet{hirashita10} show that shattering driven by
interstellar turbulence can produce
small grains efficiently if the metallicity becomes
higher than a critical value
($\sim 0.1$--1 Z$_{\sun}$). Thus, the
critical metallicity for the grain growth is near the
metallicity level where shattering produces small
grains efficiently. The production of small grains
accelerates the grain growth by accretion, which
raises the grain abundance.
With the increased dust abundance,
shattering can
be further efficient, although the
produced small grains by shattering does
not necessarily activate further shattering
of large grains
\citep{jones96}. Such an interplay between
shattering and accretion
may be interesting to investigate in the future.

The size distribution of grains produced by AGB
stars is also important since AGB stars become the
dominant dust production source after several
hundreds of Myr \citep{valiante09,gall11,asano11}.
The size of grains produced in AGB stars is
suggested to be large
($a\sim 0.1~\micron$) from the observations of spectral
energy distributions \citep{groenewegen97,gauger99},
although \citet{hofmann01} show that the grains are
not single-sized. To clarify the grain size distribution
formed by AGB stars is important for the efficiency of
grain growth. Shattering in the ISM may also play a
role in efficiently producing small grains even if
AGB stars only produce large grains
\citep{hirashita10b}. In this case, efficient grain
growth can occur.

It is interesting to point out that the critical
metallicity is within the metallicity range typical
of dwarf galaxies (Table \ref{tab:Zcr}).
This confirms the conclusion by \citet{asano11} that
the strong metallicity dependence of the dust-to-gas
ratio in dwarf galaxies can be explained by the
grain growth in clouds.

The grain growth is also important in some
high-redshift galaxy populations. In fact,
high-redshift quasars have solar (or more)
metallicities \citep[e.g.][]{juarez09}, which
implies that the grain growth is indeed governing
the dust abundance
in distant quasars
\citep{michalowski10,pipino11,asano11}
\citep[but see][]{valiante09,gall11b}.
If the abundance of small grains are enhanced
because of shattering as suggested by \citet{hirashita10},
the critical metallicity becomes lower, so that the
importance of grain growth in clouds is further
pronounced.

Qualitatively it may be predicted that the
mid-infrared emission from very small grains is
relatively suppressed
if the grain growth in clouds is activated.
However, it is hard to quantitatively predict
the galaxy-scale observational features caused by the
grain growth in clouds because it is
difficult to selectively see the clouds, where
grain growth is occurring. Since observations of
galactic spectral energy distribution inevitably
include the emission from diffuse medium, other
mechanisms modifying the grain size distribution such as
shattering and coagulation are also reflected
in the observed emission from grains. Indeed,
\citet{galliano05}
show that the grain size distribution is biased
toward smaller grains in some dwarf galaxies,
which may be interpreted as the efficient grain
processing in diffuse medium.
{They also demonstrate that there is a variety in the
grain size distribution among dwarf galaxies.}
Such a variety will be investigated in
the future with a consistent treatment of a
nonlinear combination between the small
grain production by shattering and sputtering
and the grain growth by accretion and
coagulation.

\section{Conclusion}\label{sec:conclusion}

We have formulated and investigated the grain growth
rate by
accretion in interstellar clouds. The formalism is
applicable to any grain size distribution. We have
found that the grain size distribution is really
fundamental in regulating the grain growth rate.
We have also implemented the formulation of
grain growth in individual clouds into the
chemical evolution models of entire galaxies.
{The models also treat dust supply from
stellar sources and dust destruction by
SN shocks, but we have focused particularly on the
grain growth in clouds in this paper.}
We have found that the metallicity level where the
grain growth in clouds becomes dominant strongly
depends on the grain size distribution.
If the significant fraction of the grains
have radii $\la 0.01~\micron$ or the grain
size distribution is described as power law
with $r\ga 3.5$, the large grain abundance at
the sub-solar metallicity level is naturally
explained by the grain growth in clouds because
the surface-to-volume ratio of the grains is
large enough. The grain growth should be
efficient in galaxies whose
metallicity is above $Z_\mathrm{cr}$
estimated in Section \ref{subsec:Zcr}.
Our formulation for the grain growth is applicable
to any grain size distribution and is implemented
straightforwardly into any framework of chemical
enrichment models.

\section*{Acknowledgments}
We are grateful to the referee, A. P. Jones, for useful
comments, which improved the discussion in this
paper very much.
We thank A. K. Inoue for helpful
discussions on dust evolution in galaxies.
H.H. is supported by NSC grant 99-2112-M-001-006-MY3.

\appendix

\section{Instantaneous recycling approximation}
\label{app:ejecta}

We estimate $\mathcal{R}$ and $\mathcal{Y}_{Z}$
used in Section \ref{subsec:enrichment}. With an
initial mass function (IMF) $\phi (m)$, $\mathcal{R}$ and
$\mathcal{Y}_{Z}$ are written as
\begin{eqnarray}
\mathcal{R} & = & \int_{m_t}^{m_\mathrm{u}}[m-w(m)]
\phi (m)\,\mathrm{d}m,\\
\mathcal{Y}_{Z} & = & \int_{m_t}^{m_\mathrm{u}}m
p_\mathrm{Z}(m)\phi (m)\,\mathrm{d}m,
\end{eqnarray}
where $m_t$ is the turn-off stellar mass, $m_\mathrm{u}$
is the upper mass cutoff of stellar mass, $m$ is the stellar
mass, $w_m$ is the remnant mass,
$p_\mathrm{Z}(m)$ is the fraction of mass converted into
metals in a star of mass $m$. We assume the Salpeter
IMF ($\phi (m)\propto m^{-2.35}$) with stellar mass
range
$0.1~\mathrm{M}_{\sun}\leq m\leq 100~\mathrm{M}_{\sun}$.
The IMF is normalized as
\begin{eqnarray}
\int_{m_\mathrm{l}}^{m_\mathrm{u}}m\phi (m)\,\mathrm{d}m
=1,
\end{eqnarray}
where $m_\mathrm{l}$ is the lower mass cutoff of stellar
mass.

\begin{figure}
\includegraphics[width=0.45\textwidth]{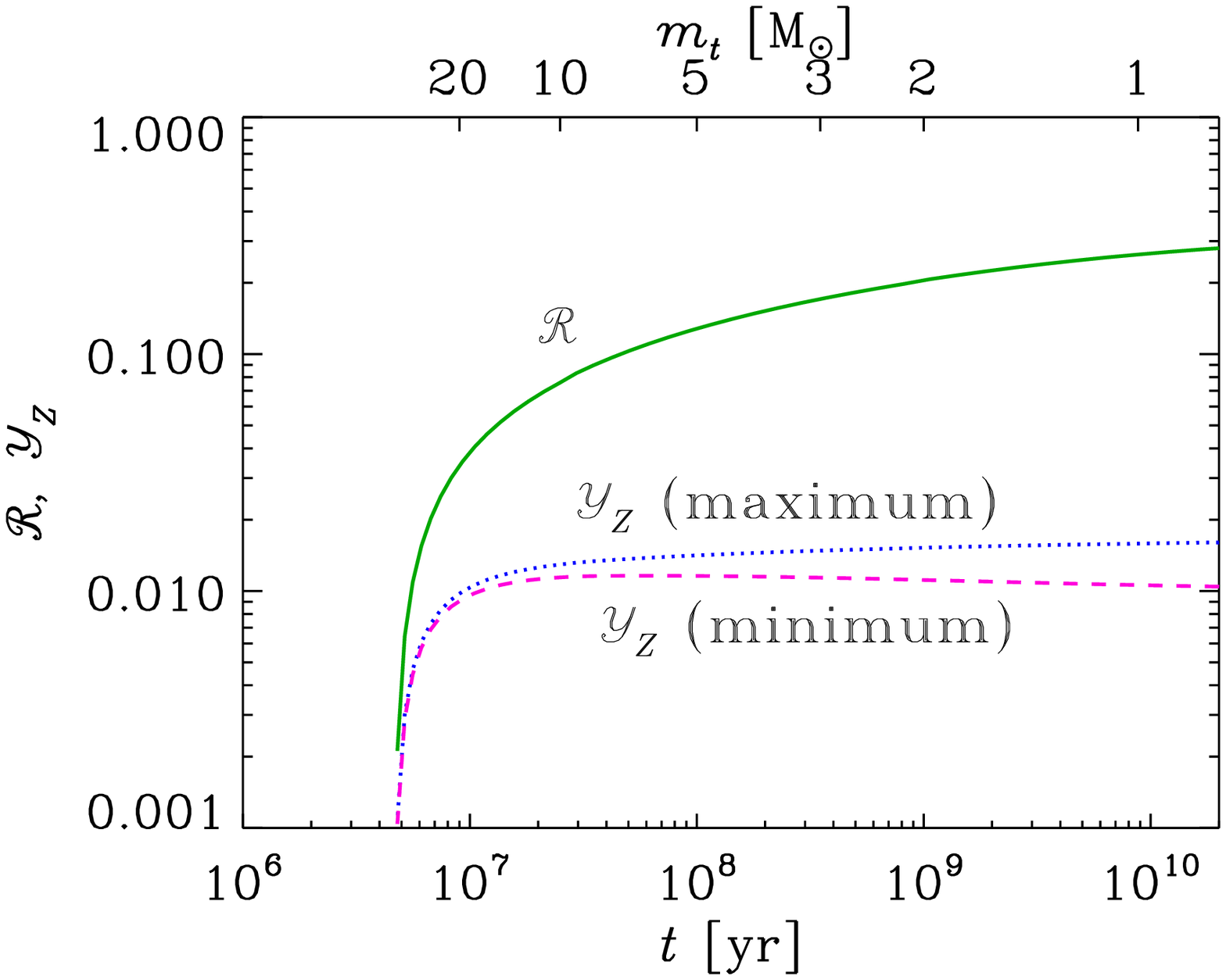}
 \caption{The returned fraction of gas $\mathcal{R}$
 (solid line) and the fraction of newly produced metals
 $\mathcal{Y}_Z$ (dotted and dashed lines) are shown
 as a function of turn-off stellar mass or age. For
 $\mathcal{Y}_Z$, the maximum (dotted line) and
 the minimum (dashed line) cases are shown.
 }
 \label{fig:returned}
\end{figure}

For the remnant mass, we adopt the fitting formula provided
by \citet{inoue11}:
\begin{eqnarray}
\frac{w(m)}{m}=\left\{
\begin{array}{ll}
1 & (m>40~\mathrm{M}_{\sun}), \\
0.13\left({\displaystyle\frac{m}{8~\mathrm{M}_{\sun}}}
\right)^{-0.5} &
(8\leq m\leq 40~\mathrm{M}_{\sun}), \\
0.13\left({\displaystyle\frac{m}{8~\mathrm{M}_{\sun}}}
\right)^{-0.7} &
(m<8~\mathrm{M}_{\sun}).
\end{array}
\right.
\end{eqnarray}
We also adopt the fitting formula for the mass of ejected
metals as a function of stellar mass as
\citep{inoue11}
\begin{eqnarray}
\frac{m_Z(m)}{m}=\left\{
\begin{array}{ll}
0 & (m>40~\mathrm{M}_{\sun}), \\
0.02\left({\displaystyle\frac{m}{8~\mathrm{M}_{\sun}}}
\right)^{2} &
(8\leq m\leq 40~\mathrm{M}_{\sun}), \\
0.02\left({\displaystyle\frac{m}{8~\mathrm{M}_{\sun}}}
\right)^{0.7} &
(m<8~\mathrm{M}_{\sun}).
\end{array}
\right.
\end{eqnarray}
\citet{inoue11} show that there is no trend with
metallicity $Z$ for $m_Z/m$. In fact, $m_Z$ is the
mass of ejected metals,
so we need to subtract the metal mass already
included before the nucleosynthesis:
$mp_Z(m)=m_Z(m)-Z[m-w(m)]$. Since the
metallicity range of interest is $0\leq Z\la 0.02$,
$mp_Z(m)$ is reasonably between
$m_Z(m)-0.02[m-w(m)]$ (called minimum) and
$m_Z(m)$ (maximum). In
Fig.\ \ref{fig:returned}, we show $\mathcal{R}$
and $\mathcal{Y}_Z$ as a function of the turn-off
mass (or age). We adopt the values at $t=5$ Gyr,
that is, $\mathcal{R}=0.18$ and
$\mathcal{Y}_Z=0.013$ (average of the maximum and the
minimum), since the
typical gas consumption time-scale or star formation
time-scale of nearby star-forming galaxies is 1--10 Gyr
\citep{kennicutt98,gavazzi02}. However, the uncertainty
caused by the age is within a factor of 2 if we adopt
an age of $>4\times 10^7$ yr. Thus, as long as we
treat galaxies whose typical star formation time-scale is
longer than a few $\times 10^7$ yr, our calculation
gives a reasonable results for the metal and dust
enrichment.

The SN rate, $\gamma$, is also necessary to estimate
the dust destruction rate. If we assume that
stars with $m\geq 8$ M$_{\sun}$ become SNe,
$\gamma$ can be related to the star formation rate
by approximating the lifetimes of these stars to be
zero:
\begin{eqnarray}
\gamma =\psi\int_{8\,\mathrm{M}_{\sun}}^{m_\mathrm{u}}
\phi (m)\,\mathrm{d}m=0.00742(\mathrm{M}_{\sun}^{-1})
\psi .\label{eq:snr}
\end{eqnarray}

\bsp

\label{lastpage}

\end{document}